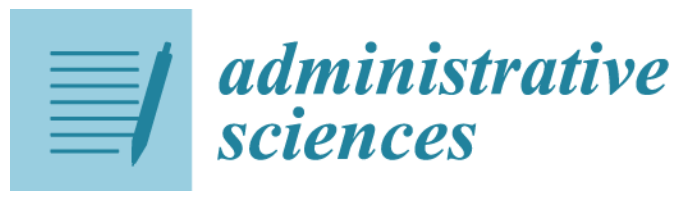

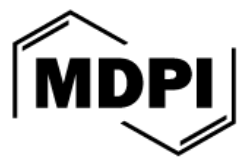

*Review*

# An Integrative Multidimensional Conceptualization of Telework Behavior: A Systematic Review and Grounded Theory Approach

**Sahar Babaei [1,*], Saeed Nosratabadi [2], Thabit Atobishi [3] and Sahar Abu Bakir [4]**

[1] Rawls College of Business, Texas Tech University, Lubbock, TX, 79409, USA; sababaei@ttu.edu
[2] Rawls College of Business, Texas Tech University, Lubbock, TX, 79409, USA; snosrata@ttu.edu
[3] Department of Health Informatics and Technology, College of Applied Medical Sciences, King Faisal University, Hofuf 31982, Saudi Arabia; tatubaishi@kfu.edu.sa
[4] Department of Business Administration, Amman Arab University, Amman 11953, Jordan; sahar@aau.edu.jo
* Correspondence: sababaei@ttu.edu

**Abstract**

Telework has expanded rapidly, and understanding the behaviors employees enact under it has become correspondingly important. This study develops an integrative conceptual framework specifying the multidimensional nature of telework behavior. A systematic literature review following PRISMA identified 114 review articles, which were analyzed using constructivist grounded theory. Six behavioral dimensions were identified, covering performance, communication, environmental, task, policy, and well-being conduct. Antecedents were grouped into individual factors, job characteristics, organizational norms, technological factors, and work environment factors. Outcomes were grouped into job satisfaction, productivity, turnover, health and well-being, work–life balance, and social isolation. Three contextual moderators were identified, namely telework modality, telework preference, and cultural and national context. Fourteen propositions link these categories and are advanced as testable claims rather than established findings, since they are derived from published review evidence and have not been empirically tested. The framework proposes telework behavior as the construct through which the conditions of remote work are translated into employee and organizational outcomes, and it specifies a content domain from which measures of the construct can be developed. It also indicates where organizations can direct policy, training, and intervention.

## 1. Introduction

Telework refers to work arrangements where employees complete job tasks away from the conventional workplace, typically relying on information and communication technologies (ICTs) to interact with colleagues, managers, and clients (Ferrara et al., 2022; Urien, 2023). The concept encompasses decentralized models such as home-based telecommuting, mobile work, remote work, and virtual work, unified by three defining features, namely work performed outside the centralized office, reliance on ICTs for communication and coordination, and altered spatial and temporal boundaries between work and personal life (Chan et al., 2023). Conceptualizations of telework have evolved considerably, shifting from a narrow focus on location and technology toward a more multifaceted perspective that captures its implications for individual, organizational, and societal outcomes, including social relationships, work–life balance, job attitudes, performance, health, and well-being (Marques et al., 2025). Growing attention has also turned to conditions that shape telework's effects, such as job characteristics, work environment, and personal dispositions (Urien, 2023), and to variations in telework frequency, autonomy, and integration as key dimensions shaping individual and organizational outcomes beyond binary telework/non-telework categorizations (Wang et al., 2023). Throughout this article, a telework arrangement refers to the formal configuration of remote work an employer establishes, comprising its location, its extent, and the conditions attached to it, as distinct from the behaviors employees enact within it.

Telework behavior refers to the actions, practices, and conduct employees exhibit while carrying out occupational tasks away from the conventional workplace (Wang et al., 2023). Scholars have examined discrete behavioral dimensions of remote work, including time use, social interactions, communication, boundary management, and performance (Podolsky et al., 2022), such as how teleworkers structure schedules, collaborate remotely, maintain connectivity, segment or integrate work and personal roles, network, and manage impressions (Athanasiadou & Theriou, 2021). However, this literature remains scattered, with research typically focusing narrowly on singular behaviors rather than telework behavior holistically, and no clear consensus on its definition or core dimensions (Athanasiadou & Theriou, 2021). Scholars have called for further theorizing to elucidate the antecedents, characteristics, and outcomes of telework behavior, as well as the contextual and individual factors that shape behavioral responses (Biron et al., 2023). Hunton and Harmon's (2004) Telework Behavior Model and Ahmad et al.'s (2022) constraints–coping–effectiveness framework offer partial conceptualizations linking telework conditions, employee behaviors, and outcomes, yet neither comprehensively captures the full multidimensional nature of telework behavior across its performance, communication, environmental, task, policy, and well-being facets. Synthesis is further challenged by the diversity of disciplines, theories, and methodologies used to study discrete aspects of teleworker behavior, underscoring the need for an integrative conceptualization that consolidates findings across disciplines to advance research and theory on telework behavior.

Given this lack of integration, the present study aims to address this gap by developing a conceptual framework that synthesizes the current literature and delineates the key dimensions of telework behavior, defined here as the actions, practices, and conduct exhibited by teleworkers as they manage role boundaries, communicate, collaborate, and complete occupational tasks remotely (Podolsky et al., 2022). Grounded in this conceptualization, the study addresses the following research question.

RQ: What are the core dimensions of telework behavior and how are they related to employee and organizational outcomes?

By addressing this question, the study develops an integrative conceptual model that specifies telework behavior together with the antecedents that precede it, the outcomes that follow from it, and the conditions under which these relationships vary. This study offers three theoretical contributions. First, it integrates antecedents, enacted behaviors, outcomes, and contextual moderators into a single nomological network for telework behavior, whereas prior reviews address these segments separately and examine single behaviors such as communication or boundary management in isolation (Cunha et al., 2024; Tapasco-Alzate et al., 2024). Second, this architecture positions behavior as the construct standing between telework conditions and their effects, where existing reviews link the telework arrangement directly to satisfaction, productivity, and turnover without specifying what employees do in between (Marques et al., 2025; Onnis et al., 2026). Third, it conceptualizes telework behavior as a multidimensional construct with a proposed content domain of six dimensions, a precondition for developing and validating the measures of telework behavior that the field currently lacks.

## 2. Materials and Methods

The overarching goal of this study was twofold: to systematically synthesize current literature related to telework and to leverage these research findings to develop a review-derived conceptual model that theorizes telework behaviors. To fulfill the first aim of aggregating evidence on telework arrangements and their outcomes, a systematic literature review methodology following the PRISMA (Preferred Reporting Items for Systematic Reviews and Meta-Analyses) guidelines was adopted. Systematic reviews provide a rigorous, transparent, and reproducible means of identifying, selecting, appraising, and synthesizing research evidence related to a particular topic or phenomenon of interest (Page et al., 2021). Using systematic search and screening procedures, literature reviews minimize bias and maximize coverage of available evidence, providing a solid empirical foundation to inform theory building.

To fulfil the second aim of integrating the systematic review conclusions into a coherent conceptual framework, this study employed a constructivist grounded theory approach (Charmaz, 2014). Grounded theory provides an apt qualitative methodology geared towards inductively generating theoretical models grounded in empirical observations. It offers systematic guidelines for coding textual data from literature to define concepts, categories, and relationships to explain dynamic phenomena like telework behavior. Further, constructivist grounded theory acknowledges the interpretative role of the researchers in conceptualizing the data through an iterative process of theme definition and integration around a unifying theoretical framework. The constructivist orientation was thus aligned with this study's objective of developing an explanatory theory of telework behavior that is substantiated by yet extends beyond the collated literature evidence.

Alternative qualitative approaches were considered. Thematic analysis identifies recurring patterns across a body of text but stops at describing those patterns and does not require specifying how categories relate to one another, which would not have produced the linkages between antecedents, behaviors, and outcomes that this study sought (Lambert, 2019). Qualitative content analysis typically applies a coding frame defined in advance and reports the frequency with which categories occur, an orientation that presumes the categories are already known. Framework synthesis codes evidence against an existing conceptual framework, which presupposes the framework that this study set out to construct. Grounded theory was selected because it is the approach designed to build categories and their relationships from the data rather than to describe or classify against a scheme fixed beforehand, and among its variants the constructivist version was chosen because it treats the resulting model as a construction shaped by the researchers rather than as a discovery of an objective structure (Charmaz, 2014; Morse et al., 2021).

### *2.1. Systematic Literature Review*

A systematic literature search was conducted following the PRISMA guidelines (Page et al., 2021) to identify relevant review articles on telework. The search was first performed in January 2026 and updated in June 2026 using the SCOPUS database, which contains over 79 million records across scientific and social science disciplines. As outlined in the Identification phase of PRISMA (Figure 1), the search strategy involved applying keywords including "Telework" OR "Remote Work" OR "Home Office" OR "Work from Home" OR "Hybrid Work" OR "Mobile Work" OR "Virtual Work". These were searched in article titles, abstracts, and keyword fields in order to maximize identification of appropriate review articles. The initial search yielded 25,081 records. Results were subsequently limited to review-type articles ($n$ = 789) and further restricted to English-language publications ($n$ = 746). As shown in the Screening phase of the PRISMA flow diagram (Figure 1), the titles and abstracts of these 746 records were screened for relevance. The screening criteria required that records represent literature reviews directly related to the topic of telework. Application of these criteria led to the exclusion of 516 records, leaving 230 records to continue to full-text review. Next, in the Eligibility phase, full-text copies of the remaining 230 records underwent in-depth assessment based on predetermined eligibility criteria. To qualify for inclusion, articles had to represent a literature review focused on evaluating different aspects of telework including adoption, implementation, management, effects, outcomes, etc. After review, 116 records did not meet the eligibility criteria and were excluded. The reason for exclusion was recorded for each record at the full-text stage: telework was not the primary focus of the article, in that remote work was addressed only incidentally within a broader topic ($n$ = 57); the topic fell outside the scope of the review question ($n$ = 52); the report duplicated or overlapped with another review already included ($n$ = 4); and the full text could not be retrieved ($n$ = 3). The full list of excluded reports and the reason for each is provided in the Supplementary material. The decision to restrict inclusion to review articles followed from the objective of the study. The aim was not to estimate the magnitude of any particular telework relationship, which would require primary studies reporting comparable effects, but to identify the behavioral constructs that the field has established across a large and dispersed literature. Review articles are suited to this purpose because each one already aggregates findings from many primary studies, so the 114 included articles provide coverage across disciplines, occupations, and national settings that screening primary studies at a comparable scale would not have permitted. Inclusion in a review also functions as an evidence threshold, in that a construct appearing across multiple independent reviews has accumulated sufficient primary evidence to warrant synthesis rather than resting on a single study. Empirical primary studies were therefore excluded as a unit of analysis, although their findings enter the corpus indirectly through the reviews that synthesize them.

Screening was conducted by three authors. At the title and abstract stage, all three independently screened a random 15% subsample of the 746 records ($n$ = 112) against the inclusion criteria, and agreement was almost perfect (Fleiss' $\kappa$ = 0.81). At the full text stage, all three independently assessed a random 15% subsample of the 230 records ($n$ = 35), yielding substantial agreement (Fleiss' $\kappa$ = 0.77). The remaining records were divided among the three coders, with disagreements resolved by discussion. The final result, as depicted in the Included phase of Figure 1, was 114 eligible literature review articles to be included for data extraction and synthesis (see Table A1 in Appendix A). The search covered publications from 1981 through June 2026, and the search, screening, and eligibility assessment were carried out over six months between January 2026 and June 2026. No formal risk-of-bias, reporting-bias, or certainty assessment was conducted.

### *2.2. Constructivist Grounded Theory*

This study adopts a constructivist grounded theory approach (Charmaz, 2014) to analyze the literature review findings and develop a conceptual model theorizing telework behavior. Grounded theory emphasizes developing theoretical explanations grounded in iterative analysis of empirical data, providing systematic guidelines for coding textual data, defining analytic categories and themes, and integrating them into a theoretical framework (Birks & Mills, 2015). Unlike objectivist grounded theory, the constructivist variant acknowledges that analysis is shaped by researchers' prior perspectives rather than producing an exact representation of objective reality, aiming instead to construct reputable, original, and significant contributions by remaining closely entrenched in the data throughout theory building (Charmaz, 2014). This approach was chosen for its suitability in developing context-based, process-oriented explanations of complex concepts, allowing rigorous development of an integrated conceptual framework theorizing the processes shaping telework behaviors (Charmaz, 2014).

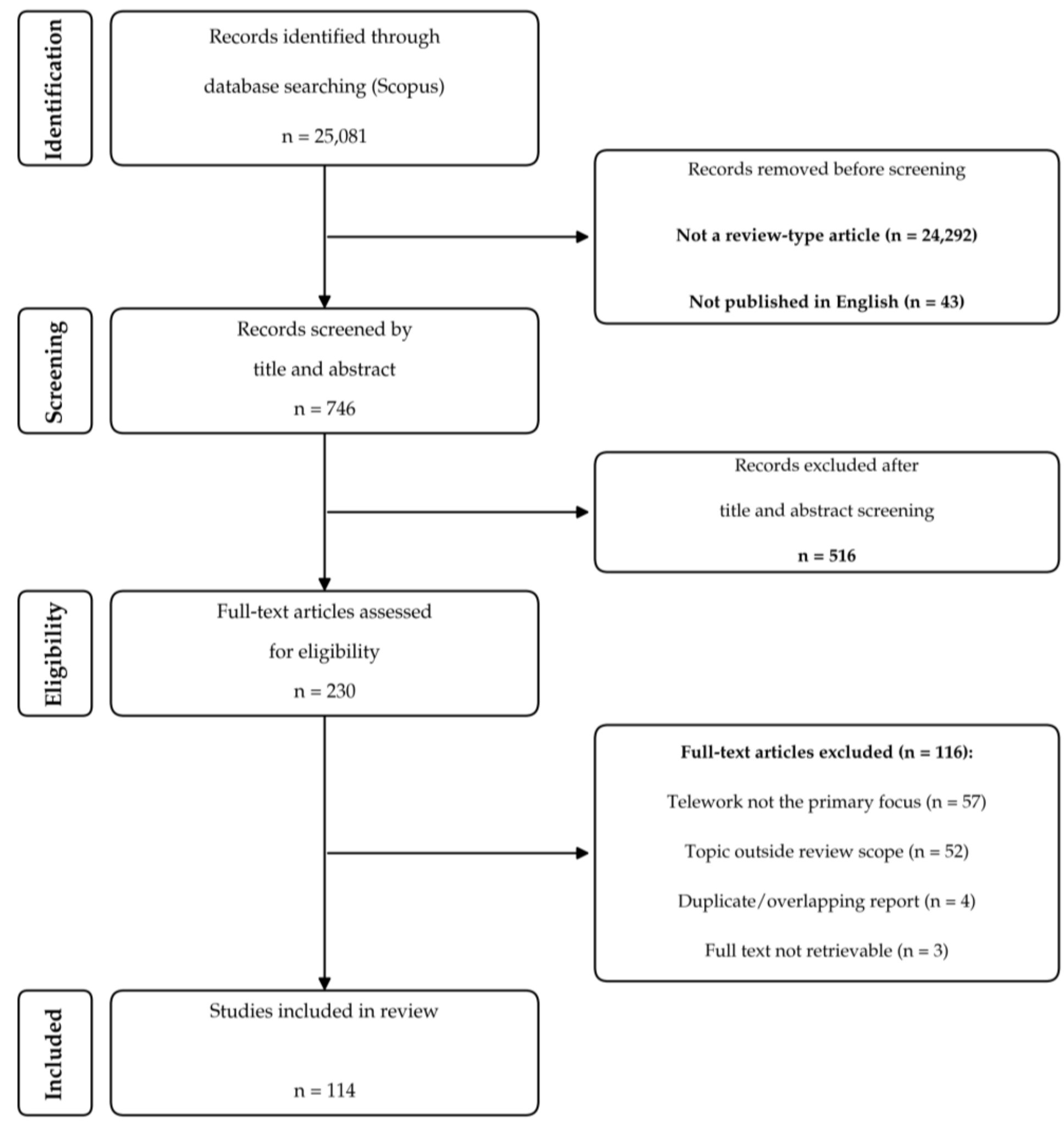

**Figure 1.** PRISMA flow diagram depicting literature screening and selection process.

Grounded theory was also selected because it provides an established and widely used set of analytic protocols. Together with PRISMA, which governs how the evidence base was identified, screened, and selected, it makes each step of the research process explicit and reproducible, so the resulting framework can be traced back to the evidence from which it was built. Applying grounded theory to review articles rather than to interview or observational data does not conflict with the method. Charmaz (2014) treats existing texts as a legitimate data source, and the rigor of grounded theory rests on its coding procedures, constant comparison, and theoretical integration rather than on the type of data analyzed. The requirement is that categories are constructed from the data rather than imposed from a preexisting theoretical scheme, and this holds for published texts as it does for interview transcripts. Grounded theory has been adapted for this purpose in the analysis of published literature (Wolfswinkel et al., 2013) and in the synthesis of secondary qualitative evidence into theoretical models (Hoon, 2013).

Grounded theory also fits the objective of this study. The research question concerns the dimensional structure of telework behavior as it appears across a fragmented literature, so the appropriate unit of analysis is the accumulated body of review evidence rather than a single sample of teleworkers. Each of the 114 included reviews synthesizes findings from many primary studies, which gives the analysis coverage across disciplines, occupations, and countries that primary data collection could not provide. Grounded theory provides systematic procedures for moving from this dispersed textual evidence to defined categories and their linkages, which is what conceptualizing telework behavior required.

Following Charmaz's (2014) guidelines, the analysis proceeded through three rounds. The first round applied line-by-line initial coding to the findings and discussion sections of the 114 articles, defining actions and processes described in each and producing 43 initial codes. Coding was conducted in MAXQDA 24, and constant comparison was applied by testing each new excerpt against the excerpts already assigned to a code, which surfaced the boundary problems that the codebook revision addressed. Comparison across codes at this stage produced the classification criteria reported in Section 3.1, distinguishing enacted actions from the conditions preceding them, the states resulting from them, and the settings conditioning them. Applying those criteria led to five relabelings, in which codes naming a condition rather than an action were renamed to name the action, and to the separation of organizational culture from cultural and national context.

The second round applied focused coding to the codes retained as behaviors, grouping them by the target of the action into the six dimensions reported in Section 3.2, and grouping the antecedent, outcome, and moderator codes by conceptual similarity into their respective categories. Categories were compared against one another to test whether each dimension was distinct, which is what led ergonomic practices and posture management to be placed with workspace-directed behaviors rather than with health-directed ones. The third round integrated the categories around telework behavior as the core category, selected because every other category in the analysis stood in relation to it either as a precursor, a result, or a condition. Integration consisted of establishing which antecedent categories and which outcomes the reviews linked to each behavioral dimension, and these linkages became the propositions stated in Section 3.3. Theoretical sampling was applied within the corpus by returning to articles addressing categories that remained thin, which is how the moderator codes were identified. Memos written at each round record the reasoning behind category definitions and boundary decisions and constitute the audit trail for the analysis.

To document analytic consistency, the three coders independently applied the initial codebook of 43 codes to a random 15% subsample of coded excerpts. First round agreement was moderate (Fleiss' $\kappa = 0.51$). Boundary rules distinguishing stress, technostress, and exhaustion were then specified and a fresh subsample was recoded, yielding $\kappa = 0.83$. Assignment of the 43 initial codes to the behavior, antecedent, moderator, and outcome categories reached $\kappa = 0.75$, and assignment of the 12 behavioral codes to the six dimensions reached $\kappa = 0.80$. Coding was managed in MAXQDA 24, and Fleiss' $\kappa$ was computed externally from the exported coding matrices. Consistent with the constructivist orientation of the analysis (Charmaz, 2014), these coefficients are reported as evidence of analytic transparency rather than as a claim that the categories are independent of the researchers' interpretation. The final conceptualization was developed jointly and is supported by an audit trail of analytic memos.

## 3. Findings

### 3.1. Initial Coding

The systematic literature search and screening process resulted in 114 eligible articles for inclusion and data extraction (Figure 1). These articles underwent thorough review to identify initial codes—concepts describing factors related to telework behaviors—as part of the constructivist grounded theory approach. Example initial codes included work–life balance, social isolation, job performance, and communication. In total, 43 distinct initial codes were identified from relevant excerpts across the articles. These initial codes and sample literature sources are listed in Table 1, representing the first phase of grounded theory analysis to capture the breadth of telework factors described across the reviews.

**Table 1.** Initial codes and example source articles from phase 1 coding.

| Initial Codes | Sources |
|---|---|
| Work–life balance | Beckel and Fisher (2022), Ferrara et al. (2022), Crawford (2022), Furuya et al. (2022), De Vincenzi et al. (2022), Chan et al. (2023), Verma et al. (2023), Bhat et al. (2023) |
| Social isolation | Beckel and Fisher (2022), Ferrara et al. (2022), Miyake et al. (2022), Nguyen (2021), Craig et al. (2022), Gualano et al. (2023) |
| Job performance | Ferrara et al. (2022), Mutiganda et al. (2022), Herrera et al. (2022), Anakpo et al. (2023) |
| Job satisfaction | Ferrara et al. (2022), Miyake et al. (2022), Efimov et al. (2022), Craig et al. (2022), Chow et al. (2022) |
| Productivity | Ferrara et al. (2022), Crawford (2022), De Vincenzi et al. (2022), Mutiganda et al. (2022), Anakpo et al. (2023) |
| Stress | Beckel and Fisher (2022), De Vincenzi et al. (2022), Dávila Morán et al. (2023), Dávila Morán (2023) |
| Communication | Nguyen (2021), Arunprasad et al. (2022), Lundqvist and Wallo (2023) |
| Trust | Nguyen (2021), Arunprasad et al. (2022), Seo and Kim (2023) |
| Engagement | Nguyen (2021), Efimov et al. (2023) |
| Technology adoption | Espitia et al. (2022), Gagné et al. (2022) |
| Collaboration | Šímová and Zychová (2023), Willox et al. (2023) |
| Training | Šímová and Zychová (2023), Seo and Kim (2023) |
| Organizational culture | Šímová and Zychová (2023), Seo and Kim (2023), Chan et al. (2023) |
| Job demands | Miyake et al. (2022) |
| Technological support | De Vincenzi et al. (2022), Bahamondes-Rosado et al. (2023) |
| Time management | Allen et al. (2015), Gajendran and Harrison (2007) |
| Focus | Fonner and Roloff (2010), Pyöriä (2011) |
| Boundary management | Allen et al. (2015), Kossek et al. (2006) |
| Organizational policy adherence | Hunton and Harmon (2004), Bailey and Kurland (2002) |
| Motivation | Gagné et al. (2022) |
| Job design | Gagné et al. (2022), Crawford (2022) |
| Leadership | Efimov et al. (2022), Lundqvist and Wallo (2023) |
| Mental health management | Efimov et al. (2022), Dávila Morán et al. (2023), Santurtún and Shaman (2023) |
| Noise control | Ang and Cui (2022) |
| Ergonomic practices | Beckel and Fisher (2022), Camboim et al. (2023) |
| Physical health management | Wells et al. (2023), Santurtún and Shaman (2023) |
| Psychological health | Wells et al. (2023) |
| Work–life flow | Wells et al. (2023) |
| Technostress | Ferrara et al. (2022), Bahamondes-Rosado et al. (2023) |
| Musculoskeletal Disorder (MSD) | Wodajeneh et al. (2023) |
| Personality | Parra et al. (2022), Gavoille and Hazans (2022) |
| Psychosocial factors | Antunes et al. (2023) |
| Exhaustion | Bhat et al. (2023) |
| Meaningfulness | Palumbo et al. (2023) |
| Well-being | Morán et al. (2022), Anakpo et al. (2023) |
| Workload | Morán et al. (2022) |
| Turnover | Mutiganda et al. (2022) |
| Physical work environment | Herrera et al. (2022), Santurtún and Shaman (2023) |
| Posture management | Camboim et al. (2023) |
| Back pain | Camboim et al. (2023) |
| Telework modality | Antunes et al. (2023), Verma et al. (2023) |
| Telework preference | Urien (2023) |
| Cultural and national context | Khanjani et al. (2026), Yousef (2026) |

Upon reviewing the role of the elements identified in initial codes in the original articles, it became evident that these elements could be classified into four overarching categories, namely telework behavior factors, antecedents of telework behavior, outcomes of telework behavior, and moderators, the contextual conditions under which these relationships vary. To conceptualize the dimensions of telework behavior itself, it was necessary to distinguish the behavioral factors from their precursors and results, and explicit criteria were applied for this purpose. A factor was classified as a telework behavior when it denotes an observable action or practice enacted by the teleworker and under the teleworker's volitional control. A factor was classified as an antecedent when it denotes a condition, resource, or disposition that precedes and shapes such action but is not itself enacted by the teleworker. A factor was classified as an outcome when it denotes a psychological, physical, or organizational state that results from telework behavior. A factor was classified as a moderator when the reviews report that it alters the strength or direction of a relationship between other factors rather than acting as a precursor or a result in its own right. Applying these criteria, codes describing experienced states rather than enacted actions, such as stress, exhaustion, technostress, and social isolation, were classified as outcomes, and codes describing given conditions of the remote work setting were classified as antecedents. Where a code named both a condition and the action taken to manage it, the code was relabeled to name the action, so that noise and posture appear as noise control and posture management. The moderator criterion distinguishes cultural and national context from culture, in that the former refers to the societal and country-level setting in which telework is enacted and is reported to condition how telework relates to its outcomes, whereas the latter refers to the shared assumptions and norms of the employing organization and operates as an antecedent. Therefore, the initial list of factors was further analyzed and organized into these four categories, which are delineated in Table 2.

Categorizing the factors in this manner enabled a conceptual delineation of telework behavior and its nomological network. The 43 initial codes were distributed as 12 telework behaviors, 13 antecedents, 15 outcomes, and 3 moderators. The telework behavior factors represent the key dimensions of telework behavior itself. These are the observable actions, practices, and conduct exhibited by employees as they work remotely. The antecedent factors are those that shape or predict telework behavior, while the outcomes are the results emanating from telework behavior. The moderators are the conditions under which the reviews report these relationships to differ in strength or direction. Organizing the factors into these categories provides conceptual clarity regarding the nature of telework behavior and its place within a broader framework. This classification formed the basis for developing a multidimensional definition and model of telework behavior.

**Table 2.** Categorization of initial codes into telework behavior, antecedents, outcomes, and moderators.

| Telework Behaviors | Antecedents | Outcomes | Moderators |
|---|---|---|---|
| Job performance | Organizational culture | Job satisfaction | Telework modality |
| Communication | Technological support | Productivity | Telework preference |

| Collaboration | Leadership | Turnover | Cultural and national context |
|---|---|---|---|
| Time management | Job design | Well-being | |
| Focus | Trust | Psychological health | |
| Boundary management | Job demands | Stress | |
| Ergonomic practices | Personality | Exhaustion | |
| Posture management | Workload | Technostress | |
| Noise control | Motivation | Social isolation | |
| Organizational policy adherence | Technology adoption | Work–life balance | |
| Mental health management | Training | Work–life flow | |
| Physical health management | Physical work environment | Meaningfulness | |
| | Psychosocial factors | Engagement | |
| | | Musculoskeletal Disorder (MSD) | |
| | | Back pain | |

### *3.2. Focused Coding*

Inspired by prior research examining behavioral dimensions of telework such as time use, social interactions, communication, boundary management, and performance (Bailey & Kurland, 2002; Fonner & Roloff, 2010), the aim is to conceptualize and dimensionalize the telework behavior factors identified in the literature review. In the focused coding phase of constructivist grounded theory, factors identified in the previous phase and presented in Table 2 were categorized based on their similarity into main dimensions and labeled accordingly. Six focused codes were identified for the factors classified as Telework Behavior in the first column of Table 2. These are (1) Performance Behavior, which encompasses job performance factors; (2) Communication Behavior, which includes communication and collaboration factors; (3) Environmental Behavior, which incorporates ergonomic practices, posture management, and noise control factors; (4) Task Behavior, including time management, focus, and boundary management; (5) Policy Behavior, including organizational policy adherence; and (6) Well-Being Behavior, which comprises mental health management and physical health management. This dimensionalization provides an initial conceptual overview of the key types of actions and conduct that comprise telework behavior, as indicated by the literature. The six categories are proposed as distinct but interrelated behavioral dimensions that teleworkers exhibit and enact while working remotely. See Figure 2 for the initial conceptual dimensions of telework behavior.

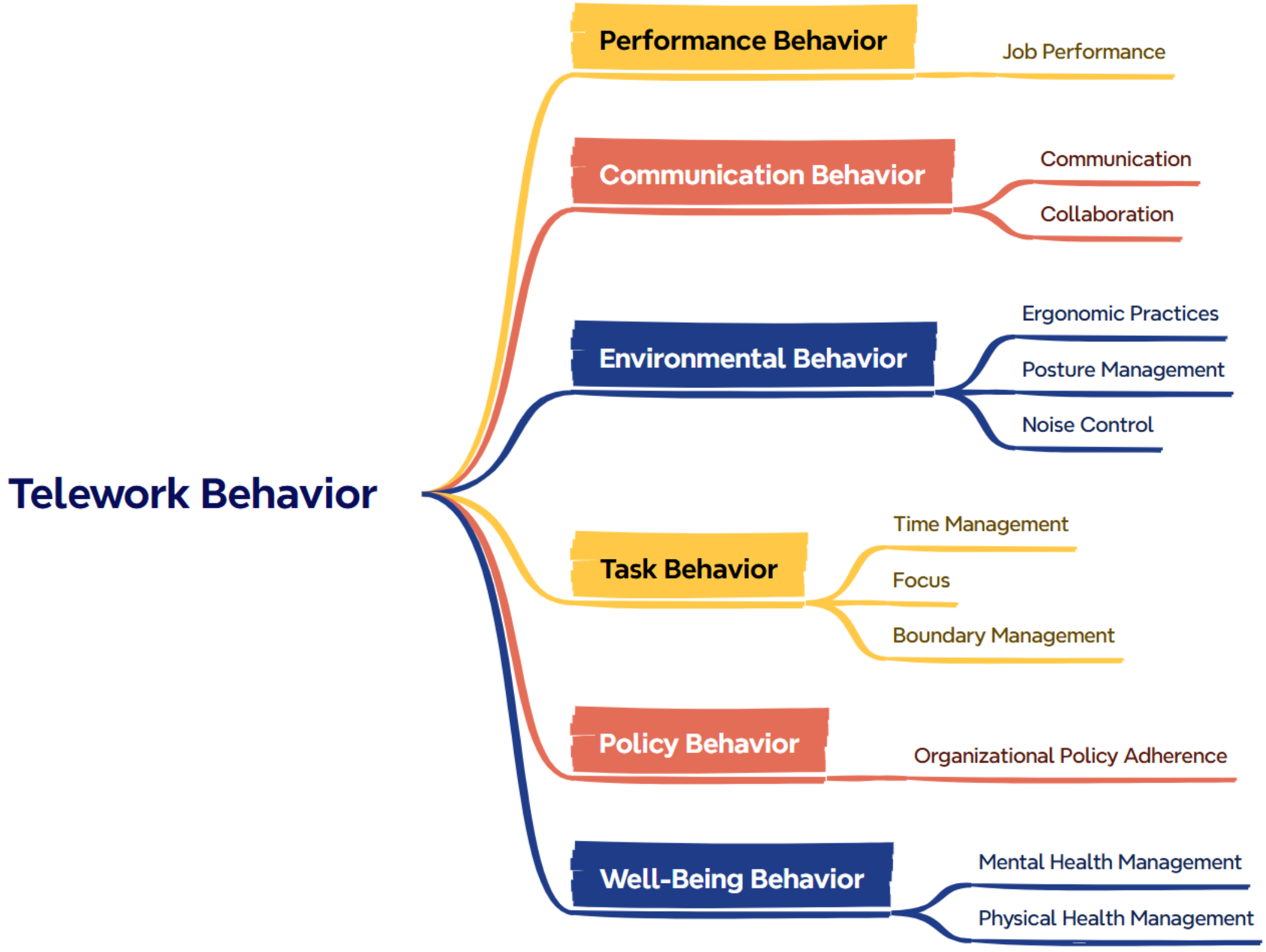

**Figure 2.** Focused codes for conceptual dimensions of telework behavior.

Five focused codes were identified for the factors identified as antecedents in the previous phase and presented in Table 2. These are (1) Individual Factors, including personality and motivation; (2) Job Characteristics, encompassing workload, job demands, and job design factors; (3) Organizational Norms, including culture, trust, leadership, and training factors; (4) Technological Factors, including technological support and technology adoption; and (5) Work Environment Factors, encompassing the physical work environment and psychosocial factors of the remote setting. See Figure 3 for identified antecedents of telework behavior.

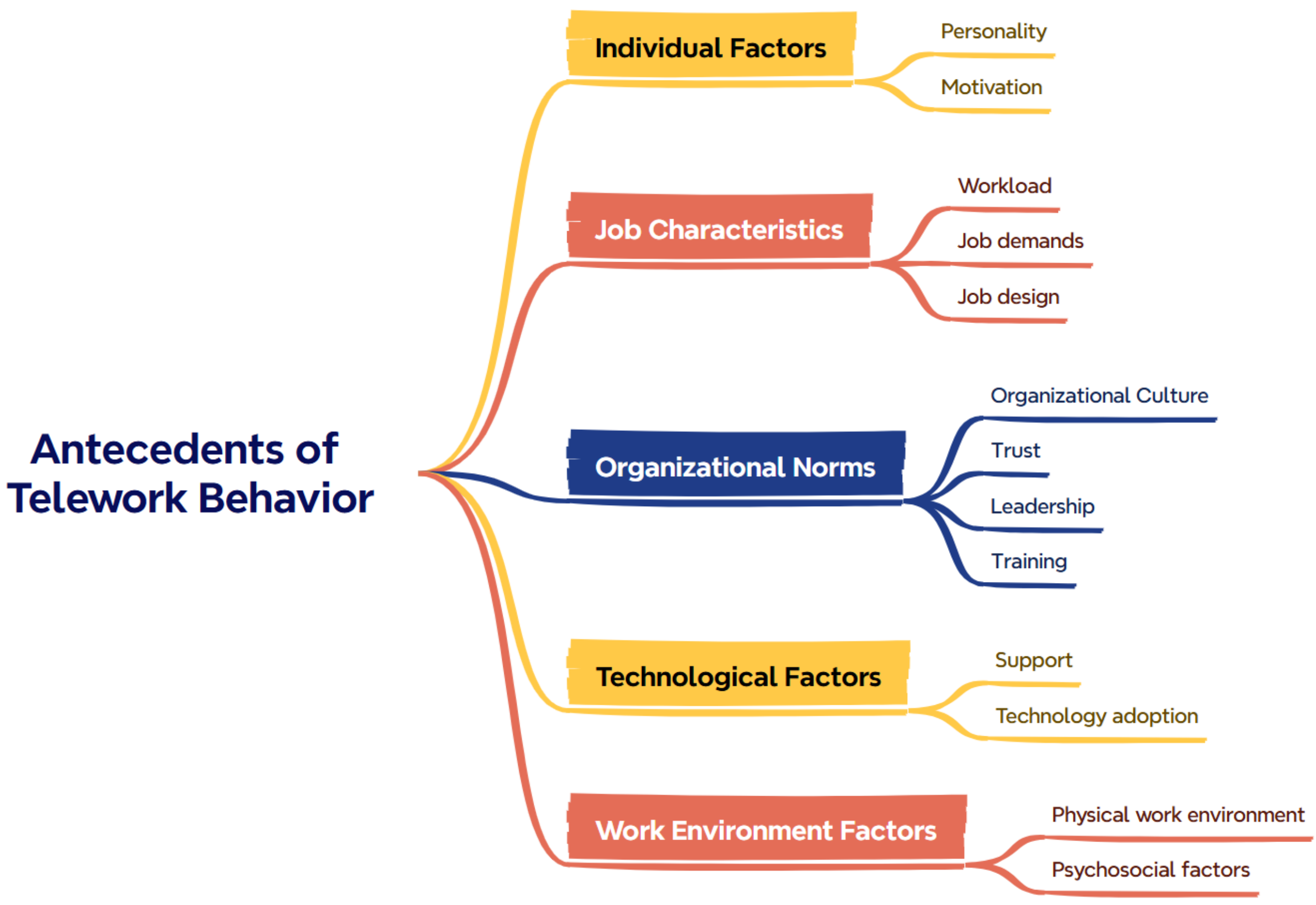

**Figure 3.** Focused codes for antecedents of telework behavior.

Six focused codes were identified for the factors classified as outcomes in the previous phase and presented in Table 2. These are (1) Job Satisfaction, encompassing job satisfaction, engagement, and meaningfulness as attitudinal outcomes; (2) Productivity; (3) Turnover; (4) Health and Well-Being, encompassing well-being, psychological health, stress, exhaustion, technostress, musculoskeletal disorder, and back pain; (5) Work–Life Balance, encompassing work–life balance and work–life flow; and (6) Social Isolation. Job satisfaction, productivity, and turnover were the most frequently identified outcomes across the reviewed literature and are retained as distinct focused codes on that basis, while the remaining outcome codes were grouped according to their conceptual similarity. See Figure 4 for the focused codes for telework behavior outcomes.

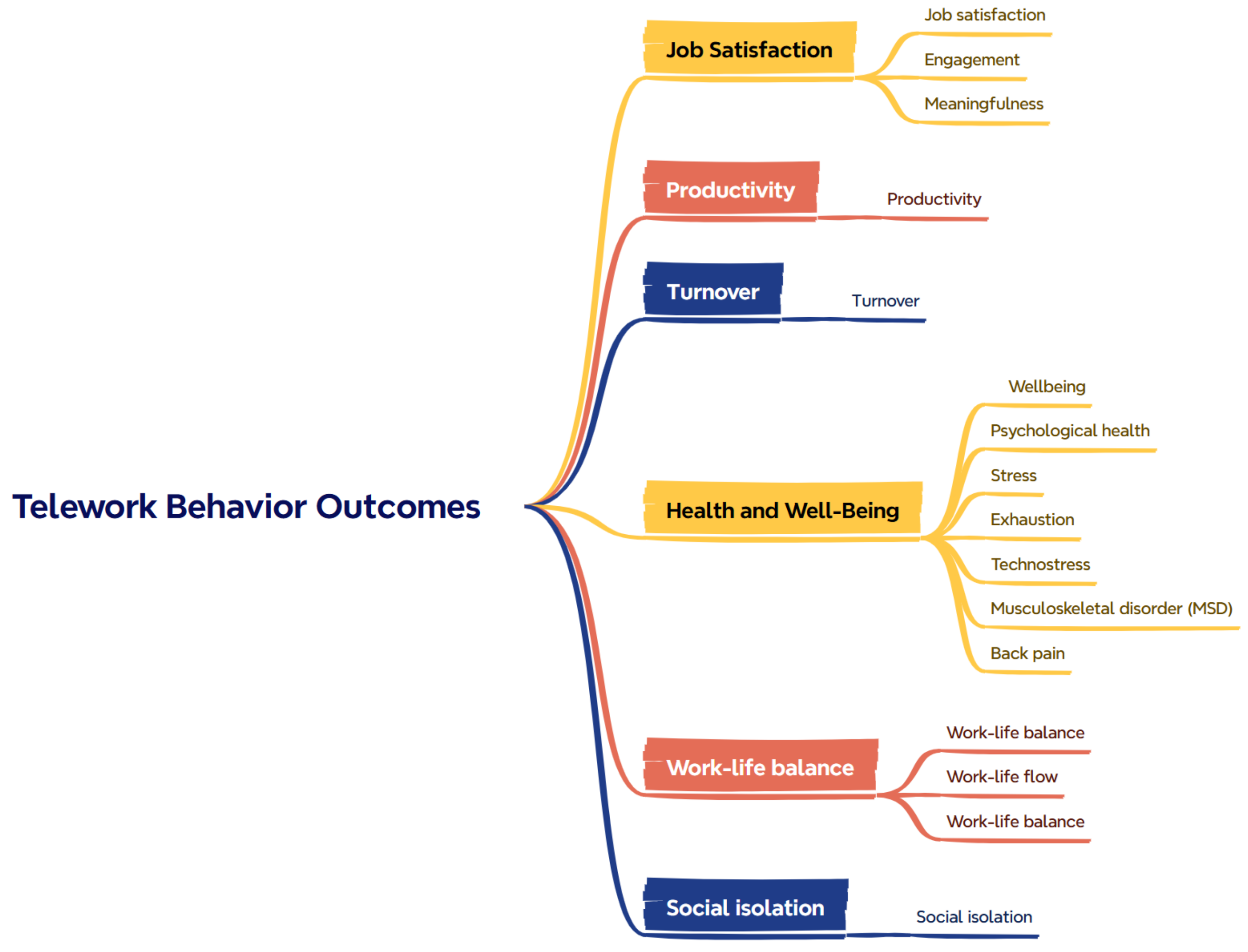

**Figure 4.** Focused codes for telework behavior outcomes.

One focused code, Contextual Moderators, was identified for the factors classified as moderators in the previous phase and presented in Table 2. It comprises three elements. Telework modality refers to the extent of the arrangement, distinguishing part-time from full-time telework, and reviews comparing these arrangements report that the psychosocial risks and benefits associated with telework differ between them (Antunes et al., 2023; Verma et al., 2023). Telework preference refers to the degree to which the arrangement matches what the employee wants, which reviews report to condition how telework relates to well-being (Urien, 2023). Cultural and national context refers to the societal and country-level setting in which telework is enacted, which reviews report to shape how remote arrangements operate and what they produce across national settings (Khanjani et al., 2026; Yousef, 2026). These three elements are grouped as a single focused code because each specifies a condition under which the relationships among antecedents, behaviors, and outcomes are reported to vary, rather than a precursor or a result in its own right (see Figure 5).

**Figure 5.** Focused code for contextual moderators of telework behavior.

### *3.3. Conceptualization*

The categories developed through focused coding are integrated here into a nomological network for telework behavior, meaning a specification of the construct together with the antecedents that precede it, the outcomes that follow from it, and the conditions under which these relationships vary (Cronbach & Meehl, 1955). This integration proceeds in four steps. The behavioral dimensions are defined first, since they constitute the construct being conceptualized. The antecedent categories are then specified, and propositions linking them to the behavioral dimensions are derived. The outcome categories follow, with propositions linking the behavioral dimensions to them. The contextual moderators are specified last. Telework behavior occupies the intervening position in this network, so the antecedent and outcome propositions together define it as the construct through which the conditions of remote work are translated into employee and organizational results. The propositions state relationships inferred from the reviewed literature and are advanced as testable claims rather than as established findings.

Telework involves a complex array of actions and conduct exhibited by employees as they complete occupational tasks away from the traditional physical workplace. Based on an extensive review of literature across multiple disciplines, this study delineates six key dimensions that conceptually encapsulate the diverse behavioral facets inherent in telework: performance behavior, communication behavior, environmental behavior, task behavior, policy behavior, and well-being behavior. Each dimension encompasses distinct yet interrelated behavioral elements that elucidate the skills, practices, and strategies enacted by teleworkers across an array of situations and contexts (See Table 3). Each element is assigned to a single dimension, with the assignment determined by the target of the action. Environmental behavior comprises actions directed at the physical workspace, whereas well-being behavior comprises actions directed at the teleworker's own health. The physical and psychological states that result from these actions, such as musculoskeletal complaints or stress, are treated as outcomes rather than as behavioral elements.

**Table 3.** Conceptual dimensions and elements of telework behavior.

| Dimensions | Elements | Description |
| --- | --- | --- |
| Performance Behavior | Job Performance | Setting and monitoring performance targets and directing effort toward the execution of assigned job tasks. |
| Communication Behavior | Communication | Selecting communication channels, how often teleworkers initiate exchanges, and the timeliness with which they reply to others. |
| | Collaboration | Working jointly with others on shared tasks and establishing positive interpersonal connections and rapport virtually. |
| Environmental Behavior | Ergonomic Practices | Behaviors involved in arranging equipment, seating, and workspace layout to maximize comfort, safety, and health. |
| | Posture Management | Behaviors involved in maintaining and adjusting working posture to prevent physical strain. |
| | Noise Control | Efforts to regulate distractions like noise, interruptions, and competing home demands when working remotely. |
| Task Behavior | Time Management | Behaviors used to structure one's schedule, allocate time across responsibilities, and avoid procrastination when working remotely. |
| | Focus | Concentrating on work tasks and avoiding unnecessary distractions in the telework environment. |
| | Boundary Management | Behaviors used to demarcate work and personal roles and activities, either by segmenting them or allowing flexibility across roles. |
| Policy Behavior | Organizational Policy Adherence | Following the organization's established policies, protocols, expectations, and codes of conduct relating to telework. |
| Well-Being Behavior | Mental Health Management | Behaviors involved in maintaining psychological and emotional wellness, such as maintaining social connections, engaging interests, and practicing relaxation. |
| | Physical Health Management | Behaviors involved in meeting basic self-care needs, such as nutrition, sleep, scheduled breaks, and regular exercise. |

Performance behavior encompasses the job performance actions that enable teleworkers to successfully complete job tasks remotely. Teleworkers set and monitor their own targets and indicators in the absence of co-location with managers, and direct effort toward assigned duties under the autonomy and flexibility telework affords (Efimov et al., 2022).

Communication behavior encompasses the communication and collaboration through which teleworkers exchange information and work jointly with managers and colleagues, a vital dimension given the lack of in-person contact and reliance on technology (Gagné et al., 2022). Skilled use of appropriate communication media enhances effectiveness and rapport-building remotely (Willox et al., 2023), while regular interaction and timely responsiveness support performance, knowledge sharing, and coordination, with their absence raising risks of isolation and collaboration breakdowns. Beyond frequency and timeliness, how teleworkers communicate and build relationships virtually also carries broader implications for social connectivity and overall telework effectiveness (Espitia et al., 2022).

Environmental behavior encompasses the actions teleworkers take to manage and optimize their physical remote workspace, including ergonomic practices, posture management, and noise control, behaviors that are essential since teleworkers must establish an effective work context themselves rather than relying on employer-managed office facilities. Ergonomic practices involve arranging seating, monitor position, lighting, and workspace layout, and these actions help prevent musculoskeletal issues, underscoring the need for ergonomics training and resources (Wodajeneh et al., 2023). Posture management, meaning the maintenance and adjustment of working posture during the workday, serves the same preventive function (Camboim et al., 2023). Teleworkers further control noise through behaviors like using headphones, setting boundaries with others at home, and maintaining dedicated workspace (Ang & Cui, 2022). What distinguishes these behaviors from well-being behavior is their target, as each is directed at configuring the workspace rather than at the teleworker's own health routines.

Task behavior involves actions and conduct that enable teleworkers to effectively structure, prioritize, and manage their job responsibilities and assigned duties in the remote work context. It encompasses time management, focus, and boundary management behaviors exhibited by teleworkers as they organize and carry out work without direct supervision. Time management refers to behaviors used to structure one's schedule, allocate time across responsibilities, and avoid procrastination when working remotely. Tracking time usage provides insight into time management behaviors (Hackney et al., 2022; Tapasco-Alzate et al., 2024). Focus involves concentrating on work tasks and avoiding unnecessary distractions in the telework environment (Keating et al., 2024). With increased autonomy yet reduced oversight, behaviors that cultivate focus, such as removing or ignoring distractions and designating workspace, are vital for productivity (Bergefurt et al., 2023; Cunha et al., 2024). Multi-tasking and attention can be measured as indicators of focus. Boundary management means behaviors used to demarcate work and personal roles and activities, either by segmenting them or allowing flexibility across roles. Teleworkers must consciously manage boundaries given the blending of home and work spheres (Kossek et al., 2006). Understanding boundary preferences and strategies provides insight into this area.

Policy behavior encompasses teleworkers' adherence to organizational rules, protocols, and expectations governing remote work, along with their broader commitment to the job and organization, behaviors that organizations establish to ensure telework effectiveness from a risk and control standpoint (Hunton & Harmon, 2004; Bailey & Kurland, 2002). Given the risks of shirking and reduced oversight inherent in telework arrangements, the reviews associate clear policies around accountability, availability, communication, conduct, and work outputs with compliance, while commitment behaviors such as dependability and integrity remain important despite the oversight challenges telework poses.

Well-being behavior encompasses the actions teleworkers take to maintain and promote their own mental and physical health, a key dimension given extensive evidence of telework's beneficial and detrimental impacts on worker health outcomes (Furuya et al., 2022; Beckel & Fisher, 2022). Mental health management behaviors include maintaining social connections, engaging interests, getting outdoors, exercising, and practicing relaxation (Chow et al., 2022), while physical health management behaviors involve proper nutrition, adequate sleep, scheduled breaks, and regular exercise to meet self-care needs. These behaviors are directed at the person rather than at the workspace, which distinguishes them from the ergonomic and posture management behaviors classified under environmental behavior. The health states that follow from both sets of behaviors, such as stress, exhaustion, and musculoskeletal complaints, are treated as outcomes.

Together, these six multidimensional categories and their respective elements offer a conceptual framework for examining the behaviors exhibited by teleworkers, integrating fragmented insights from the literature into a cohesive structure for analysis. This review-based conceptualization lays a foundation for developing measures, advancing theory, and guiding future research on the relationships between teleworker behavior and individual, group, and organizational outcomes.

As shown in Figure 3, the antecedents identified from the literature were categorized into individual factors, job characteristics, organizational norms, technological factors, and work environment factors. Drawing from the literature, this study delineates these five categories as follows: individual factors (personality, motivation), job characteristics (workload, job demands, job design), organizational norms (organizational culture, trust, leadership, training), technological factors (technological support, technology adoption), and work environment factors (physical work environment, psychosocial factors). Table 4 presents these antecedent categories along with their constituent elements. For each category, a proposition is derived specifying the behavioral dimensions it is expected to shape, based on the linkages reported across the reviewed literature.

**Table 4.** Antecedent categories and their elements.

| Antecedent Categories | Elements | Description |
|---|---|---|
| Individual Factors | Personality | Personality traits and dispositions of individuals such as extraversion, conscientiousness, self-efficacy. |
| | Motivation | The individual's drive and willingness to engage in telework and its associated tasks. |
| Job Characteristics | Workload | The volume of work assigned to the teleworker and the pace at which it must be completed. |
| | Job Demands | The cognitive, emotional, and physical effort the job requires of the teleworker. |
| | Job Design | The way tasks are structured and allocated, including the independence and discretion available in work scheduling and methods. |
| Organizational Norms | Organizational Culture | Shared assumptions, values, and norms of the employing organization that shape behaviors. |
| | Trust | The degree of confidence between managers and teleworkers in each other's competence and integrity. |
| | Leadership | The conduct of managers in directing, supporting, and monitoring teleworkers at a distance. |
| | Training | Organizational provision of instruction and development opportunities for telework-related competencies. |
| Technological Factors | Technological Support | Provision of technology infrastructure, assistance, and troubleshooting. |
| | Technology Adoption | The uptake and use of the technological tools and infrastructure that telework requires. |
| Work Environment Factors | Physical Work Environment | Conditions of the remote workspace such as thermal comfort, air quality, and lighting. |
| | Psychosocial Factors | Social and organizational conditions of the remote setting such as job insecurity, role clarity, and autonomy. |

Individual factors encompass traits and dispositions inherent to the teleworker that influence behavioral responses, including personality (extraversion, conscientiousness, openness, self-efficacy, proactivity), which leads individuals to react distinctly to telework scenarios (Parra et al., 2022; Gavoille & Hazans, 2022), and motivation, which shapes the drive and willingness with which employees engage in telework tasks (Gagné et al., 2022). Because telework removes the structure that co-location provides, the self-regulatory dispositions captured by these factors are expected to bear most directly on the behaviors through which teleworkers organize their own work and manage their own health.

**Proposition 1.** *Individual factors are associated with the enactment of task behavior and well-being behavior.*

Job characteristics constitute workplace-related precursors of telework behavior, including job design, which allows customized behaviors to manage telework (Gagné et al., 2022; Crawford, 2022); workload, which sets the volume of work to be organized remotely (Morán et al., 2022); and job demands, which shape the effort directed at achieving objectives (Miyake et al., 2022). These characteristics define the volume of work and the discretion available to teleworkers over how work is scheduled and executed, and are therefore expected to bear on the behaviors through which work is organized and performed.

**Proposition 2.** *Job characteristics are associated with the enactment of task behavior and performance behavior.*

Organizational norms further shape behavior through organizational culture, which encourages or hinders adaptive telework behaviors (Arunprasad et al., 2022; Wodajeneh et al., 2023), trust, which elicits behaviors adhering to expectations (Nguyen, , 2021), leadership, which directs and supports teleworkers at a distance (Efimov et al., 2022; Lundqvist & Wallo, 2023), and training, which equips employees with the competencies telework requires (Šímová & Zychová, 2023; Seo & Kim, 2023). Because these factors establish and communicate what the organization expects of teleworkers and prepare them to meet those expectations, they are expected to bear most directly on compliance-related and performance-related conduct.

**Proposition 3.** *Organizational norms are associated with the enactment of policy behavior and performance behavior.*

Technological antecedents, including technological support and technology adoption, are reported to facilitate or constrain effective telework behaviors regardless of other precursors (Bahamondes-Rosado et al., 2023; De Vincenzi et al., 2022; Onnis et al., 2026). Since remote interaction and task execution are mediated by technology, the adequacy of this provision and the extent of its uptake are expected to condition how teleworkers exchange information with colleagues and carry out their assigned work (Espitia et al., 2022; Gagné et al., 2022).

**Proposition 4.** *Technological factors are associated with the enactment of communication behavior and performance behavior.*

Finally, work environment factors, comprising the physical conditions of the remote workspace such as thermal comfort, air quality, and lighting, as well as the psychosocial factors of the remote setting, establish the setting within which telework behaviors are enacted (Herrera et al., 2022; Manu et al., 2024; Navas-Martín et al., 2026; Antunes et al., 2023). These conditions are given features of the remote setting rather than actions taken by the teleworker, which is what distinguishes them from the environmental behaviors described above, and they define the demands to which workspace-directed and health-directed behaviors respond (Bergefurt et al., 2023).

**Proposition 5.** *Work environment factors are associated with the enactment of environmental behavior and well-being behavior.*

Together, these individual, job, organizational, technological, and environmental factors constitute the broader situational context that is proposed to give rise to telework behaviors, providing a framework for predicting and interpreting observed behaviors based on salient precursors.

This study considers job satisfaction, productivity, turnover, health and well-being, work–life balance, and social isolation as key outcomes of telework behavior. Job satisfaction reflects employees' affective and cognitive attitudes toward their job; telework can improve satisfaction through greater autonomy, flexibility, and work–life balance, though reduced social interaction and isolation can diminish it (Ferrara et al., 2022; Chow et al., 2022; Miyake et al., 2022). Productivity, measured through quantity, quality, and timeliness, tends to increase with telework depending on job fit, though it can suffer from home distractions (Anakpo et al., 2023; De Vincenzi et al., 2022; Crawford, 2022; Ferrara et al., 2022). Turnover, the rate at which employees voluntarily leave, may be reduced by improved work–life fit and flexibility, though disconnection from the organization can also enable attrition (Mutiganda et al., 2022; Efimov et al., 2022). Health and well-being outcomes encompass the psychological and physical states arising from telework, including stress, exhaustion, technostress, and musculoskeletal complaints (Beckel & Fisher, 2022; Bahamondes-Rosado et al., 2023; Gualano et al., 2023; Wodajeneh et al., 2023). Work–life balance reflects the degree of fit between work and personal domains, which telework can enhance through flexibility or erode through blurred boundaries (Bhat et al., 2023; Wells et al., 2023). Social isolation refers to the perceived lack of social connection and belonging that can accompany reduced in-person contact (Beckel & Fisher, 2022; Efimov et al., 2022). Table 5 shows the outcomes of telework behavior.

**Table 5.** Outcomes of telework behavior.

| Outcomes | Description |
| --- | --- |
| Job Satisfaction | An employee's affective and cognitive attitudes towards their job, including engagement and perceived meaningfulness of work. |
| Productivity | Employee output and efficiency in generating goods or services. |
| Turnover | The rate at which employees voluntarily leave an organization. |
| Health and Well-Being | Psychological and physical states resulting from telework, including well-being, psychological health, stress, exhaustion, technostress, and musculoskeletal complaints. |
| Work–Life Balance | The degree of fit and equilibrium between work and personal life domains, including work–life flow. |
| Social Isolation | The perceived lack of social connection, belonging, and interaction with colleagues. |

The reviewed literature links each behavioral dimension to particular outcomes rather than to all of them equally, and propositions are derived accordingly. Performance behavior concerns the setting of targets and the execution of assigned work, and the reviews associate these behaviors with output and with employees' attitudes toward their jobs (Anakpo et al., 2023; Hackney et al., 2022).

**Proposition 6.** *Performance behavior is associated with productivity and job satisfaction.*

Communication behavior concerns how much contact teleworkers maintain with colleagues and managers, and reduced or ineffective communication is repeatedly associated with perceived disconnection and with diminished job attitudes (Fonner & Roloff, 2010; Efimov et al., 2022).

**Proposition 7.** *Communication behavior is associated with social isolation and job satisfaction.*

Environmental behavior configures the physical conditions under which remote work is carried out, and the reviews link these actions to physical health complaints and to the capacity to work effectively (Wodajeneh et al., 2023; Camboim et al., 2023; Ang & Cui, 2022; Bergefurt et al., 2023).

**Proposition 8.** *Environmental behavior is associated with health and well-being and with productivity.*

Task behavior concerns how teleworkers allocate time, sustain focus, and demarcate work from personal roles, which the reviews associate with the fit between domains and with work output (Kossek et al., 2006; Hackney et al., 2022; Tapasco-Alzate et al., 2024).

**Proposition 9.** *Task behavior is associated with work–life balance and productivity.*

Policy behavior reflects adherence to organizational expectations under conditions of reduced oversight, which the reviews associate with organizational performance indicators and with employees' continued attachment to the organization (Hunton & Harmon, 2004; Mutiganda et al., 2022; Keating et al., 2024).

**Proposition 10.** *Policy behavior is associated with productivity and turnover.*

Well-being behavior comprises the health-directed actions teleworkers take on their own behalf, which the reviews associate with psychological and physical health states and, through them, with job attitudes (Chow et al., 2022; Crawford, 2022; Efimov et al., 2022).

**Proposition 11.** *Well-being behavior is associated with health and well-being and with job satisfaction.*

As shown in Figure 5, the reviewed literature further indicates that these relationships do not hold uniformly across all telework arrangements. Three contextual moderators were identified, and Table 6 presents them along with their descriptions. These factors are treated as moderators rather than as antecedents or outcomes because the reviews report them as conditions that alter the strength or direction of relationships between other factors rather than as precursors or results in their own right.

**Table 6.** Contextual moderators of the telework behavior framework.

| Category | Elements | Description |
|---|---|---|
| Contextual Moderators | Telework Modality | The extent of the telework arrangement, distinguishing part-time from full-time telework. |
| | Telework Preference | The degree of correspondence between the telework arrangement and the arrangement the employee prefers. |
| | Cultural and National Context | The societal and country-level setting in which telework is enacted. |

Telework modality distinguishes part-time from full-time arrangements, and reviews comparing the two report that exposure to psychosocial risk differs between them, with full-time arrangements associated with greater emotional demand and a more blurred home and work interface than part-time arrangements (Antunes et al., 2023; Verma et al., 2023). The same behaviors are therefore expected to carry different consequences depending on how much of the working week is conducted remotely.

**Proposition 12.** *Telework modality moderates the relationships between telework behaviors and outcomes, such that the associations with health and well-being and with work–life balance differ between part-time and full-time arrangements.*

Telework preference concerns whether the arrangement matches what the employee wants, and reviews report that well-being under telework depends on the correspondence between the arrangement and the employee’s preference rather than on the arrangement alone (Urien, 2023). Behaviors enacted under an arrangement the employee did not choose are therefore expected to relate differently to outcomes than the same behaviors enacted by choice.

**Proposition 13.** *Telework preference moderates the relationships between telework behaviors and outcomes, such that the associations with job satisfaction and with health and well-being are stronger when the arrangement corresponds to the employee’s preference.*

Cultural and national context refers to the societal and country-level setting in which telework is enacted, and reviews examining determinants across national settings report that the conditions supporting remote work and the results it produces vary between them (Khanjani et al., 2026; Yousef, 2026). Both the translation of organizational conditions into behavior and the translation of behavior into outcomes are therefore expected to differ across national settings.

**Proposition 14.** *Cultural and national context moderates the relationships between antecedents and telework behaviors and between telework behaviors and outcomes.*

Table 7 summarizes the fourteen propositions and the categories they connect.

**Table 7.** Propositions derived from the conceptual framework.

| Proposition | From | To |
|---|---|---|
| P1 | Individual Factors | Task Behavior; Well-Being Behavior |
| P2 | Job Characteristics | Task Behavior; Performance Behavior |
| P3 | Organizational Norms | Policy Behavior; Performance Behavior |
| P4 | Technological Factors | Communication Behavior; Performance Behavior |
| P5 | Work Environment Factors | Environmental Behavior; Well-Being Behavior |
| P6 | Performance Behavior | Productivity; Job Satisfaction |
| P7 | Communication Behavior | Social Isolation; Job Satisfaction |
| P8 | Environmental Behavior | Health and Well-Being; Productivity |
| P9 | Task Behavior | Work–Life Balance; Productivity |
| P10 | Policy Behavior | Productivity; Turnover |
| P11 | Well-Being Behavior | Health and Well-Being; Job Satisfaction |
| P12 | Telework Modality | Moderates behaviors to outcomes |
| P13 | Telework Preference | Moderates behaviors to outcomes |
| P14 | Cultural and National Context | Moderates antecedents to behaviors and behaviors to outcomes |

The final conceptualization synthesizing the multidimensional nature of telework behavior is depicted visually in Figure 6. This conceptual model sets out how the key antecedents of telework behavior across individual, job-related, organizational, technological, and environmental categories are proposed to shape the enactment of performance, communication, environmental, task, policy, and well-being behaviors exhibited in the remote context. In turn, these behavioral dimensions are proposed to bear on critical employee and organizational outcomes, including job satisfaction, productivity, turnover, health and well-being, work–life balance, and social isolation, with telework modality, telework preference, and cultural and national context conditioning these relationships. Because the behavioral dimensions stand between the antecedents and the outcomes in this structure, the model proposes telework behavior as the construct through which the conditions of remote work are translated into their effects. The linkages depicted are derived from the relationships reported across the reviewed literature and have not been empirically tested in this study. The arrows therefore denote proposed relationships corresponding to Propositions 1 through 14, and should be read as claims awaiting validation rather than as established causal effects. The conceptual framework integrates the disparate factors identified as shaping and resulting from telework behaviors into a coherent structure, delineating the proposed relationships between the antecedents, behaviors, outcomes, and moderating conditions that comprise the nomological network of the telework experience. This conceptual integration advances theoretical understanding of the multidimensional processes and linkages characterizing telework behavior.

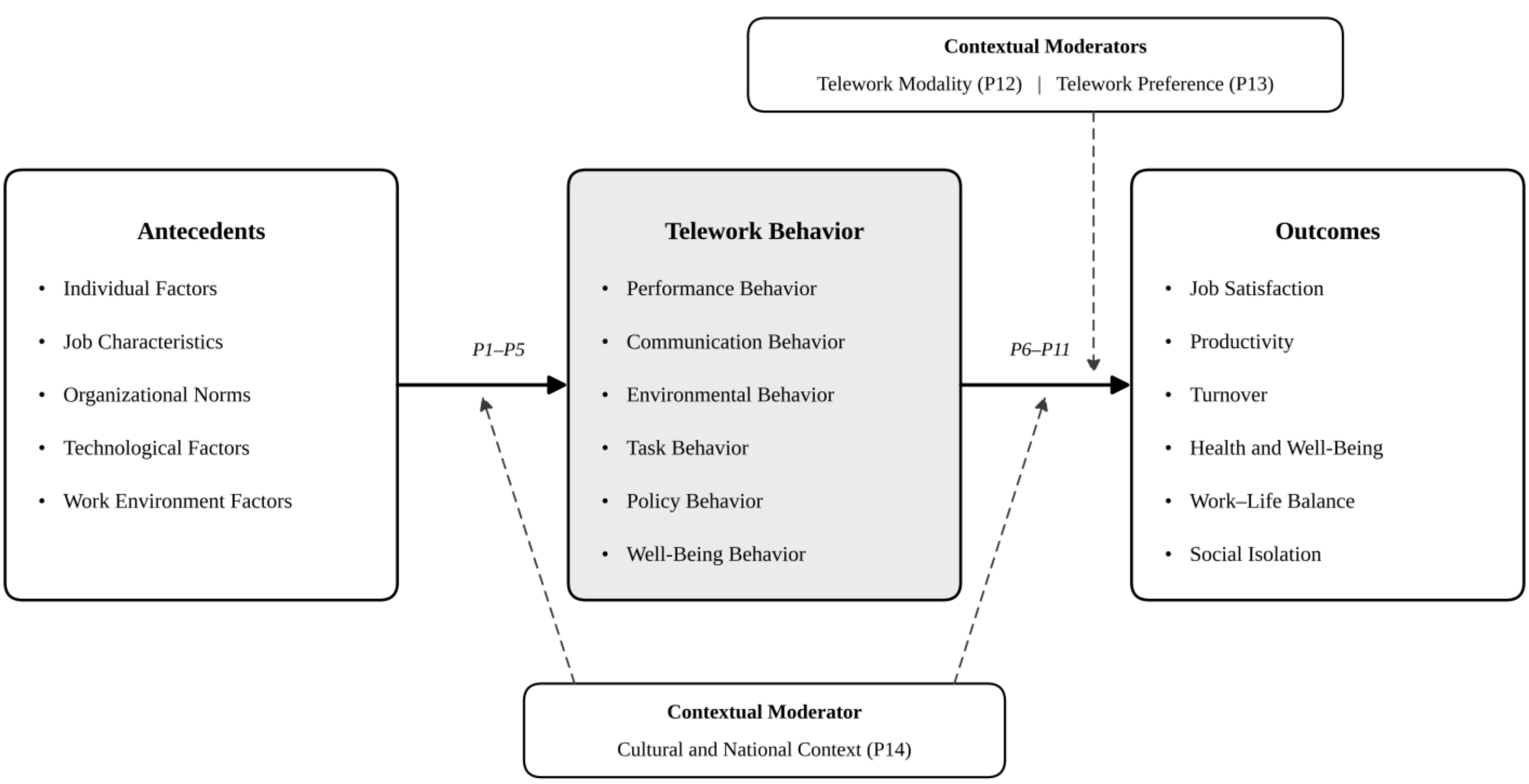

**Figure 6.** Conceptual model of telework behavior dimensions, antecedents, moderators, and outcomes.

## 4. Discussion

This study sets out to address the fragmentation that has long characterized research on telework behavior by developing an integrative conceptual model grounded in a systematic review of 114 literature reviews and a constructivist grounded theory analysis. The resulting framework specifies six behavioral dimensions, which are performance, communication, environmental, task, policy, and well-being behavior, within a nomological network of five antecedent categories, six outcomes, and three contextual moderators. It is this architecture, rather than the enumeration of behavioral dimensions, that distinguishes the framework from prior work examining isolated dimensions of telework. Rather than treating telework behavior as a single construct or as a set of loosely connected, discretely studied behaviors, the model proposes that these behavioral facets are conceptually distinct yet interrelated, each shaped by a common pool of antecedents and each linked to shared organizational outcomes.

These findings extend prior conceptualizations, but the alternatives they displace are not superseded on every count. Hunton and Harmon's (2004) Telework Behavior Model and Ahmad et al.'s (2022) constraints–coping–effectiveness framework each specify fewer constructs and tighter chains of relationships than the propositional network advanced here, which makes them more tractable to test in a single study and better suited to the specific settings for which they were developed. Bailey and Kurland's (2002) and Fonner and Roloff's (2010) treatments of communication and boundary management resolve those behaviors in more detail than the corresponding dimensions in Table 3, since a framework that covers six dimensions cannot elaborate any one of them to the depth a dedicated review achieves. The contribution claimed here is integration into a nomological network, and that gain comes at the cost of resolution within each dimension.

A more fundamental alternative concerns what occupies the position between telework conditions and their effects. An established line of work locates psychological states there rather than behaviors, treating perceived autonomy, relationship quality, and need satisfaction as the mechanisms through which telework produces its outcomes (Gajendran & Harrison, 2007; Gagné et al., 2022; Biron et al., 2023). That account is well supported and is not refuted by the present framework. The two differ in what they take to be the tractable object of intervention, since psychological states are inferred from self-report and are altered only indirectly, whereas the behaviors specified in Table 3 are observable and can be trained, scheduled, and resourced. The accounts are also compatible, in that behaviors may operate on outcomes partly through the psychological states these studies identify. Adjudicating between them requires designs that measure behaviors and psychological states in the same sample and test their relative contribution, which the review-level evidence analyzed here cannot support.

The six-dimensional structure itself is one defensible partition rather than the only one. The literature contains organizing schemes that cut the same material differently, including treatments that make boundary work the central construct from which other telework behaviors follow (Kossek et al., 2006), demands and resources framings that classify telework features by their psychosocial risk profile rather than by the actions employees take (Antunes et al., 2023), and accounts organized around accessibility and control rather than around behavioral content (Cunha et al., 2024). Consistent with the constructivist orientation of the analysis, the six dimensions are advanced as a construction shaped by the coding decisions reported in Section 3 rather than as a structure recovered from the data (Charmaz, 2014). What would test the partition is measurement. If items written from the elements in Table 3 fail to separate into six factors, or if dimensions prove empirically redundant, the structure requires revision, and the framework is stated in a form that makes that outcome detectable.

### *4.1. Theoretical Implications*

The model carries several implications for research on telework behavior. The theoretical value of the framework lies in its architecture. By distinguishing antecedents, enacted behaviors, outcomes, and contextual moderators, it treats the six dimensions as candidate mediators between telework conditions and outcomes rather than as descriptive categories, which makes the linkages empirically testable, though the review-level evidence analyzed here establishes neither mediation nor causal transmission. It also proposes a content domain for telework behavior through those dimensions, which is a requirement for developing and validating measures of the construct. Researchers can utilize the dimensions as an organizing framework to systematically investigate and compare findings within each behavioral facet as well as explore interrelationships between dimensions.

The five antecedent categories give researchers a defined set of precursors against which to predict and interpret telework behaviors. Researchers can draw from this multidimensional set of behavioral drivers to develop contextualized models suited to their specific focus. The framework also specifies three conditions under which these relationships are proposed to vary, which gives moderator research a defined starting point rather than an open field. For instance, emerging research suggests that remote work intensity relates differently to work engagement and workaholism depending on employees' psychological need satisfaction and proactive work redesign tendencies (Costantini et al., 2025), illustrating how the individual factors antecedent category identified in this framework warrants nuanced empirical attention in future studies.

The six outcome categories specify the dependent variables at issue when telework behavior is studied. Theorizing can be advanced by testing linkages between specific telework behavior dimensions and outcomes. Finally, by placing antecedents, behavioral dimensions, outcomes, and contextual moderators within a single framework, the model allows researchers to test which antecedents predict which behaviors and which behaviors predict which outcomes, and under which conditions, questions that prior segmented work has not been positioned to address.

*4.2. Practical Implications*

The framework has implications for what organizations manage. Telework policy has typically been written around the arrangement, specifying who is eligible, how many days are permitted, and what equipment is provided. Because the model positions behavior as the construct proposed to stand between the arrangement and its effects, it directs practice toward the behaviors employees enact under the arrangement. Each of the six dimensions carries distinct implications for HR policy, leadership development, performance management, employee assessment, and organizational interventions, and Table 8 sets these out. The paragraphs that follow address each area in turn.

For HR policy, the framework converts telework guidelines from statements about location into statements about expected conduct. Policy behavior is the direct target, since adherence presupposes that expectations around availability, accountability, and conduct have been written down rather than left implicit (Hunton & Harmon, 2004; Keating et al., 2024). Communication behavior suggests that policy may need to specify channel norms and response time expectations, which otherwise vary by manager and generate uneven demands on employees. Task behavior suggests that policy could address boundary management directly through provisions on core hours and disconnection, given the blending of home and work spheres under telework (Kossek et al., 2006). Environmental behavior points to workspace standards and conditions on equipment provision, and well-being behavior points to explicit break entitlements rather than treating rest as discretionary.

For leadership development, the framework identifies which behaviors managers of remote employees are positioned to shape. Because organizational norms are proposed to operate on policy behavior and performance behavior, manager training could prioritize setting explicit goals and metrics and securing adherence without resorting to surveillance, which the reviews associate with negative work behaviors rather than compliance (Keating et al., 2024; Lundqvist & Wallo, 2023). Communication behavior suggests that managers establish the interaction cadence for their teams and model responsiveness, so development may treat communication routines as a managerial responsibility rather than an individual habit. Well-being behavior suggests that managers may legitimize or discourage health-directed conduct through their own visible practice, which gives modeling a place in leadership curricula.

For performance management, the framework separates what employees do from what results. Appraisal systems for teleworkers have often relied on activity proxies such as logged hours or online status, which measure neither behavior nor outcome. The model instead specifies performance behavior and task behavior as the enacted conduct to be appraised, and productivity as the outcome against which that conduct is evaluated, which allows the two to be assessed separately. Communication behavior supplies observable indicators for this purpose, since interaction frequency and responsiveness can be documented without monitoring keystrokes or screen time.

For employee assessment, the framework informs both selection and development diagnosis. Individual factors are proposed to shape task behavior and well-being behavior, so assessment of dispositions such as conscientiousness and self-efficacy is relevant when determining suitability for remote arrangements (Parra et al., 2022; Gavoille & Hazans, 2022). Assessment can also be directed at the behaviors themselves, since the six dimensions define what a diagnostic instrument would need to cover and thereby indicate where an individual employee requires development. Telework preference is a further assessment target, because the reviews report that well-being under telework depends on the fit between the arrangement and what the employee wants rather than on the arrangement alone (Urien, 2023).

For organizational interventions, the framework indicates which intervention addresses which dimension and against which outcome it could be evaluated. Ergonomic assessment and workspace subsidy target environmental behavior and could be evaluated against health and well-being. Communication protocol redesign targets communication behavior and could be evaluated against social isolation. Boundary management training targets task behavior and could be evaluated against work–life balance. Evaluating each intervention against the outcome the framework associates with the behavior it addresses gives a more precise test than assessing all telework initiatives against general satisfaction. The moderators further indicate that intervention design may not be uniform, since the same intervention is expected to produce different results under part-time and full-time arrangements and across national settings (Antunes et al., 2023; Khanjani et al., 2026). These implications follow from the proposed linkages in the framework rather than from tested effects, so they indicate where organizations might direct policy, training, and intervention and against what those measures should be evaluated, rather than what the measures will yield.

**Table 8.** Practical implications of the six behavioral dimensions.

| Dimension | HR Policy | Leadership Development | Performance Management | Employee Assessment | Organizational Interventions |
|---|---|---|---|---|---|
| Performance Behavior | Define goal-setting cadence and reporting requirements | Train managers to set explicit targets in place of presence monitoring | Appraise goal attainment and task completion separately from productivity outcomes | Assess self-direction and capacity to work to defined targets | Introduce structured goal-setting routines |
| Communication Behavior | Specify channel norms and response time expectations | Train managers to establish team interaction cadence and model responsiveness | Use interaction frequency and responsiveness as observable indicators | Assess virtual communication and relationship-building capability | Redesign communication protocols, evaluated against social isolation |
| Environmental Behavior | Set home workspace standards and equipment provision conditions | Train managers to recognize workspace risk without inspecting homes | Consider excluding from appraisal, since workspace conditions vary by circumstance | Assess workspace adequacy at onboarding to remote work | Provide ergonomic assessment and subsidy, evaluated against health and well-being |
| Task Behavior | Establish core hours, disconnection provisions, and scheduling discretion | Train managers to support boundary setting rather than expect constant availability | Appraise scheduling and prioritization conduct separately from output | Assess time management and boundary preferences | Provide boundary and time management training, evaluated against work–life balance |
| Policy Behavior | State availability, accountability, and conduct expectations explicitly | Train managers to secure adherence without surveillance | Appraise adherence to stated expectations rather than logged activity | Assess reliability and rule-following disposition | Clarify and communicate policy, evaluated against turnover |
| Well-Being Behavior | Establish break entitlements and health support access | Train managers to model and legitimize health-directed conduct | Consider excluding from appraisal, since health conduct is personal | Assess self-care practices only where the employee consents | Provide well-being programs, evaluated against health and well-being |

*4.3. Limitations*

Restricting the search to review articles carries implications for what the framework can claim. The analysis operates at one remove from primary data, so the categories developed here reflect how review authors have already selected and aggregated primary findings, and behaviors examined in primary studies but not covered in any review would not enter the coding. The corpus also inherits the topical emphases of the review literature, which is weighted toward the pandemic period and toward health and well-being themes, so constructs reported frequently across reviews may appear more central than their underlying empirical basis warrants. The framework consequently represents consolidated rather than emerging knowledge, and behaviors introduced recently in primary research but not yet taken up in review articles would be underrepresented in the dimensions identified here. Theoretical sampling was likewise bounded by the completed systematic search rather than driven by emerging categories, so additional data could not be sought once conceptual gaps became apparent.

The review process itself has limitations. The search covered one database, so reviews indexed only elsewhere and grey literature were missed. Eligibility was restricted to English, which under-represents non-anglophone settings. The included reviews were not quality-appraised, so evidence of varying strength enters the framework unweighted. Because the corpus consists of reviews, a single primary study may be represented in several of them. The review was neither registered nor conducted against a written protocol.

Two features of the framework follow from the nature of review-level evidence rather than from choices made in the analysis. The model specifies the dimensions of telework behavior and their linkages but not the processes through which these unfold over time, because review articles report aggregated relationships rather than situated accounts. It also specifies relationships in one direction only. The reviewed literature draws predominantly on cross-sectional evidence and does not report the reciprocal effects that would warrant specifying feedback loops, such as whether outcomes like social isolation subsequently alter the communication behaviors that produced them. Both limitations point to the same requirement. The model and its fourteen propositions are derived from published review evidence and have not been empirically tested, so they require validation against primary data, beginning with the development of measures from the elements in Table 3 and proceeding to longitudinal designs capable of examining how these behaviors develop and how outcomes feed back into them.

## 5. Conclusions

This study presented an integrative conceptual framework specifying the multidimensional nature of telework behavior, based on a systematic review of 114 literature reviews and a constructivist grounded theory analysis. The framework defines six dimensions of telework behavior and places them within a nomological network of five antecedent categories, six outcomes, and three contextual moderators, connected by fourteen propositions. The integration of these elements into a single network, rather than the identification of the behavioral dimensions alone, constitutes the study's principal contribution.

Several recommendations for telework policy and practice follow from the framework, pending validation of its propositions. Organizations may consider assessing which employees' dispositions, preferences, and jobs suit telework rather than applying a uniform arrangement. Training could target telework competencies in virtual teamwork, communication, time management, and technical proficiency. Telework policies could state expectations around availability, conduct, and performance while allowing flexibility in scheduling at the job level. Technological infrastructure and troubleshooting support are identified in the framework as antecedents of communication and performance behavior.

For future research, measures of telework behavior need to be developed from the elements specified in Table 3 and validated before the propositions can be tested. Studies can then examine which antecedents predict which behaviors and which behaviors predict which outcomes. The three moderators identified here require testing, and further moderators such as telework tenure remain open, since the reviewed literature does not address them. Longitudinal designs are needed to establish how telework behaviors develop as arrangements mature and whether outcomes feed back into the behaviors that produced them. Research is also needed on interventions targeting each behavioral dimension and on their effects on the outcomes the framework associates with them.

Telework arrangements continue to grow in prevalence, and the behaviors employees enact under them are proposed here to account for much of what those arrangements produce. This study offers a conceptual model of those behaviors and a set of propositions that future research can test.

**Supplementary Materials:** The following supporting information can be downloaded at https://doi.org/10.6084/m9.figshare.33391108.

**Author Contributions:** Conceptualization, S.B. and S.N.; methodology, S.B. and T.A.; software, S.N.; validation, S.B. and S.A.B.; formal analysis, S.N. and T.A.; investigation, S.B., T.A., and S.A.B.; writing—original draft preparation, S.B. and S.N.; writing—review and editing, S.B., S.N., T.A., and S.A.B.; visualization, S.B.; supervision, T.A. All authors have read and agreed to the published version of the manuscript.

**Funding:** This research received no external funding.

**Institutional Review Board Statement:** Not applicable. This study did not involve human participants, animals, or the collection of primary data; it is based exclusively on the analysis of previously published, publicly available literature.

**Informed Consent Statement:** Not applicable, as this study did not involve human participants.

**Data Availability Statement:** No new primary data were created in this study. The dataset analyzed consists of the 114 peer-reviewed literature review articles identified through the systematic search described in the method Section, all of which are publicly available and listed in the References section and Supplementary Materials.

**Acknowledgments:** During the preparation of this manuscript/study, the author(s) used Claude Sonnet 4.6 (Anthropic) solely for the purposes of grammar and language editing of the authors' original text. The authors have reviewed and edited the output and take full responsibility for the content of this publication.

**Conflicts of Interest:** The authors declare no conflicts of interest.

## Appendix A

Table A1. Review articles included in the systematic review (N = 114).

| No. | Authors | Year | Title | Journal / Source | DOI |
|---|---|---|---|---|---|
| 1 | Abiddin et al. | 2022 | A literature review of work from home phenomenon during COVID-19 toward employees' performance and quality of | Frontiers in Psychology | 10.3389/fpsyg.2022.819860 |

| No. | Authors | Year | Title | Journal / Source | DOI |
|---|---|---|---|---|---|
| | | | life in Malaysia and Indonesia | | |
| 2 | Adegoke et al. | 2025 | Remote work, well-being, and healthy labor force participation among older adults: a scoping review | International Journal of Environmental Research and Public Health | 10.3390/ijerph22111719 |
| 3 | Alencar et al. | 2025 | Musculoskeletal pain and risk factors in office workers versus teleworkers: a systematic review | Work | 10.1177/10519815241289675 |
| 4 | Allen et al. | 2015 | How effective is telecommuting? Assessing the status of our scientific findings | Psychological Science in the Public Interest | 10.1177/1529100615593273 |
| 5 | Anakpo et al. | 2023 | The impact of work-from-home on employee performance and productivity: a systematic review | Sustainability | 10.3390/su15054529 |
| 6 | Ang and Cui | 2022 | Remote work: aircraft noise implications, prediction, and management in the built environment | Applied Acoustics | 10.1016/j.apacoust.2022.108978 |
| 7 | Antunes et al. | 2023 | Part-time or full-time teleworking? A systematic review of the psychosocial risk factors of telework from home | Frontiers in Psychology | 10.3389/fpsyg.2023.1065593 |
| 8 | Arunprasad et al. | 2022 | Exploring the remote work challenges in the era of COVID-19 pandemic: review and application model | Benchmarking: An International Journal | 10.1108/BIJ-07-2021-0421 |
| 9 | Asadieh and Neisch | 2025 | Liberation from location ties: a descriptive systematic review of shifts in location perception during and after the COVID-19 pandemic | Transportation Research Interdisciplinary Perspectives | 10.1016/j.trip.2025.101395 |

| No. | Authors | Year | Title | Journal / Source | DOI |
|---|---|---|---|---|---|
| 10 | Bahamondes-Rosado et al. | 2023 | Technostress at work during the COVID-19 lockdown phase (2020–2021): a systematic review of the literature | Frontiers in Psychology | 10.3389/fpsyg.2023.1173425 |
| 11 | Bailey and Kurland | 2002 | A review of telework research: findings, new directions, and lessons for the study of modern work | Journal of Organizational Behavior | 10.1002/job.144 |
| 12 | Beckel and Fisher | 2022 | Telework and worker health and well-being: a review and recommendations for research and practice | International Journal of Environmental Research and Public Health | 10.3390/ijerph19073879 |
| 13 | Bergefurt et al. | 2023 | How physical home workspace characteristics affect mental health: a systematic scoping review | Work | 10.3233/WOR-220505 |
| 14 | Bhat et al. | 2023 | Revolutionizing work-life balance: unleashing the power of telecommuting on work engagement and exhaustion levels | Cogent Business and Management | 10.1080/23311975.2023.2242160 |
| 15 | Booker et al. | 2025 | Technology and its influence on teleworker well-being: a systematic review | Asia Pacific Journal of Human Resources | 10.1111/1744-7941.70019 |
| 16 | Bouaziz et al. | 2025 | Assessing telework in industrial companies: a systematic literature review | International Journal of Product Lifecycle Management | 10.1504/IJPLM.2025.150952 |
| 17 | Bozzi | 2024 | Digital nomadism from the perspective of places and mobilities: a literature review | European Transport Research Review | 10.1186/s12544-024-00663-z |
| 18 | Camboim et al. | 2023 | Posture monitoring in healthcare: a systematic mapping study and taxonomy | Medical and Biological Engineering and Computing | 10.1007/s11517-023-02826-z |

| No. | Authors | Year | Title | Journal / Source | DOI |
|---|---|---|---|---|---|
| 19 | Camp et al. | 2022 | A millennial manager skills model for the new remote work environment | Management Research Review | 10.1108/MRR-01-2021-0076 |
| 20 | Chan et al. | 2023 | Work, life and COVID-19: a rapid review and practical recommendations for the post-pandemic workplace | Asia Pacific Journal of Human Resources | 10.1111/1744-7941.12355 |
| 21 | Chew and Zainal | 2022 | Building a resilient change-oriented virtual leadership framework for the higher education sector: a narrative review | Journal of Higher Education Policy and Leadership Studies | 10.52547/johepal.3.4.135 |
| 22 | Choudhary and Jain | 2025 | A systematic literature review to explore the antecedents of employee engagement among remote workers | Journal of Work-Applied Management | 10.1108/JWAM-11-2023-0136 |
| 23 | Chow et al. | 2022 | Teleworking from home experiences during the COVID-19 pandemic among public health workers (TelEx COVID-19 study) | BMC Public Health | 10.1186/s12889-022-13104-y |
| 24 | Çivilidağ and Durmaz | 2026 | The relationship of flexible working arrangements on work-family conflict, work-life balance and organizational commitment: a systematic review and meta-analysis | BMC Psychology | 10.1186/s40359-026-04216-y |
| 25 | Contreras et al. | 2020 | E-leadership and teleworking in times of COVID-19 and beyond: what we know and where do we go | Frontiers in Psychology | 10.3389/fpsyg.2020.590271 |
| 26 | Craig et al. | 2022 | Applying restorative environments in the home office while sheltering-in-place | Human Factors | 10.1177/0018720820984286 |

| No. | Authors | Year | Title | Journal / Source | DOI |
|---|---|---|---|---|---|
| 27 | Crane et al. | 2023 | Interventions designed to support physical activity and disease prevention for working from home: a scoping review | International Journal of Environmental Research and Public Health | 10.3390/ijerph20010073 |
| 28 | Crawford | 2022 | Working from home, telework, and psychological wellbeing? A systematic review | Sustainability | 10.3390/su141911874 |
| 29 | Cruz-Ausejo and Rimache | 2022 | Complications associated with remote work during the COVID-19 pandemic: a quick review | Revista de la Facultad de Medicina Humana | 10.25176/RFMH.v22i4.4806 |
| 30 | Cunha et al. | 2024 | The axis of accessibility and the duality of control of remote workers: a literature review | Journal of Information Technology | 10.1177/02683962231208218 |
| 31 | Dávila Morán | 2023 | Influence of remote work on the work stress of workers in the context of the COVID-19 pandemic: a systematic review | Sustainability | 10.3390/su151612489 |
| 32 | de Andrade et al. | 2024 | On meetings involving remote software teams: a systematic literature review | Information and Software Technology | 10.1016/j.infsof.2024.107541 |
| 33 | de Croon et al. | 2005 | The effect of office concepts on worker health and performance: a systematic review of the literature | Ergonomics | 10.1080/00140130512331319409 |
| 34 | De Macêdo et al. | 2020 | Ergonomics and telework: a systematic review | Work | 10.3233/WOR-203224 |
| 35 | De Vincenzi et al. | 2022 | Consequences of COVID-19 on employees in remote working: challenges, risks and | International Journal of Environmental Research and Public Health | 10.3390/ijerph191811672 |

| No. | Authors | Year | Title | Journal / Source | DOI |
|---|---|---|---|---|---|
| | | | opportunities — an evidence-based literature review | | |
| 36 | Dogra and Parrey | 2024 | Work from home amid black swan event (Covid-19): a bibliometric analysis from a social science perspective | Kybernetes | 10.1108/K-09-2022-1348 |
| 37 | Ducas et al. | 2025 | The impact of telework on absenteeism, presenteeism, and return to work among workers with health conditions: a scoping review | Frontiers in Public Health | 10.3389/fpubh.2025.1655200 |
| 38 | Edvardsson and Gardarsdottir | 2023 | Navigating uncharted waters: exploring leaders' challenges in the era of COVID-19 and the rise of telework | Sustainability | 10.3390/su152316471 |
| 39 | Efimov et al. | 2022 | Virtual leadership in relation to employees' mental health, job satisfaction and perceptions of isolation: a scoping review | Frontiers in Psychology | 10.3389/fpsyg.2022.960955 |
| 40 | Efimov et al. | 2023 | Healthy leadership and workplace health promotion as a team effort: a scoping review | International Journal of Environmental Research and Public Health | 10.3390/ijerph20043693 |
| 41 | Elbaz et al. | 2022 | Teleworking and work–life balance during the COVID-19 pandemic: a scoping review | Canadian Psychology | 10.1037/cap0000330 |
| 42 | Espitia et al. | 2022 | Pandemic trade: COVID-19, remote work and global value chains | The World Economy | 10.1111/twec.13117 |
| 43 | Ferdous et al. | 2026 | An integrated framework of remote work from organizational adoption to employee outcomes: a | International Journal of Management Reviews | 10.1111/ijmr.70002 |

| No. | Authors | Year | Title | Journal / Source | DOI |
|---|---|---|---|---|---|
| | | | systematic literature review | | |
| 44 | Ferrara et al. | 2022 | Investigating the role of remote working on employees' performance and well-being: an evidence-based systematic review | International Journal of Environmental Research and Public Health | 10.3390/ijerph191912373 |
| 45 | Figueiredo et al. | 2025 | Loneliness and isolation in the era of telework: a comprehensive review of challenges for organizational success | Healthcare | 10.3390/healthcare13161943 |
| 46 | Fonner and Roloff | 2010 | Why teleworkers are more satisfied with their jobs than are office-based workers: when less contact is beneficial | Journal of Applied Communication Research | 10.1080/00909882.2010.513998 |
| 47 | Furuya et al. | 2022 | Health impacts with telework on workers: a scoping review before the COVID-19 pandemic | Frontiers in Public Health | 10.3389/fpubh.2022.981270 |
| 48 | Gagné et al. | 2022 | Understanding and shaping the future of work with self-determination theory | Nature Reviews Psychology | 10.1038/s44159-022-00056-w |
| 49 | Gajendran and Harrison | 2007 | The good, the bad, and the unknown about telecommuting: meta-analysis of psychological mediators and individual consequences | Journal of Applied Psychology | 10.1037/0021-9010.92.6.1524 |
| 50 | Gavoille and Hazans | 2022 | Personality traits, remote work and productivity | SSRN Working Paper | 10.2139/ssrn.4233436 |
| 51 | Gillespie et al. | 2025 | Remote and hybrid work in crime victim services: a scoping review | Trauma, Violence, and Abuse | 10.1177/15248380251397414 |
| 52 | Giovanis et al. | 2026 | Systematic review update of organisational-level mental health promotion | International Archives of Occupational and | 10.1007/s00420-025-02193-0 |

| No. | Authors | Year | Title | Journal / Source | DOI |
|---|---|---|---|---|---|
| | | | interventions: evidence from healthcare, construction, and telework-based mobile work settings | Environmental Health | |
| 53 | Gualano et al. | 2023 | TEleworRk-relAted stress (TERRA), psychological and physical strain of working from home during the COVID-19 pandemic: a systematic review | Workplace Health and Safety | 10.1177/21650799221119155 |
| 54 | Hackney et al. | 2022 | Working in the digital economy: a systematic review of the impact of work from home arrangements on personal and organizational performance and productivity | PLoS ONE | 10.1371/journal.pone.0274728 |
| 55 | Hajjami and Crocco | 2024 | Evolving approaches to employee engagement: comparing antecedents in remote work and traditional workplaces | European Journal of Training and Development | 10.1108/EJTD-10-2022-0103 |
| 56 | Han et al. | 2025 | Leveraging digital transformation in hybrid workplaces: current landscape, underlying challenges, and future prospects | Frontiers in Human Dynamics | 10.3389/fhumd.2025.1622625 |
| 57 | Herrera et al. | 2022 | Teleworking: the link between worker, family and company | Systems | 10.3390/systems10050134 |
| 58 | Hong et al. | 2025 | Work disability and musculoskeletal disorders among teleworkers: a scoping review | Journal of Occupational Rehabilitation | 10.1007/s10926-024-10184-0 |
| 59 | Hook et al. | 2020 | A systematic review of the energy and climate impacts of teleworking | Environmental Research Letters | 10.1088/1748-9326/ab8a84 |

| No. | Authors | Year | Title | Journal / Source | DOI |
|---|---|---|---|---|---|
| 60 | Hou and Sing | 2025 | Transformative response in office workplace: a systematic review of post-pandemic changes | Buildings | 10.3390/buildings15091519 |
| 61 | Hunton and Harmon | 2004 | A model for investigating telework in accounting | International Journal of Accounting Information Systems | 10.1016/j.accinf.2004.08.001 |
| 62 | Intan et al. | 2025 | Prevalence of low back pain among office workers during the COVID-19 pandemic in various countries: a systematic review | Kesmas: Jurnal Kesehatan Masyarakat Nasional | 10.7454/kesmas.v20i1.1391 |
| 63 | Jain et al. | 2024 | Remote working and its facilitative nuances: visualizing the intellectual structure and setting future research agenda | Management Research Review | 10.1108/MRR-01-2022-0057 |
| 64 | Kang and Oh | 2026 | Toward the 'new normal': a systematic review of remote work and implications for human resource development (HRD) | Human Resource Development Review | 10.1177/15344843261441203 |
| 65 | Keating et al. | 2023 | Virtual work conditions impact negative work behaviors via ambiguity, anonymity, and (un)accountability: an integrative review | Journal of Applied Psychology | 10.1037/apl0001126 |
| 66 | Khanjani et al. | 2026 | Work without walls: a systematic review of cultural and national determinants in hybrid and remote work | European Journal of Training and Development | 10.1108/EJTD-01-2026-0010 |
| 67 | Kim et al. | 2026 | Navigating leadership in hybrid or remote workplaces: a systematic review of employee engagement strategies | Human Resource Development Review | 10.1177/15344843251381386 |

| No. | Authors | Year | Title | Journal / Source | DOI |
|---|---|---|---|---|---|
| 68 | Kossek et al. | 2006 | Telecommuting, control, and boundary management: correlates of policy use and practice, job control, and work–family effectiveness | Journal of Vocational Behavior | 10.1016/j.jvb.2005.07.002 |
| 69 | Kumar et al. | 2024 | Bibliometric analysis of remote working: 20-year literature review | Human Resource Development Review | 10.1177/15344843241305920 |
| 70 | Lundqvist and Wallo | 2023 | Leadership and employee well-being and work performance when working from home: a systematic literature review | Scandinavian Journal of Work and Organizational Psychology | 10.16993/sjwop.199 |
| 71 | Maity and Lee | 2025 | The impact of remote and hybrid work models on small and medium-sized enterprises productivity: a systematic literature review | SN Business and Economics | 10.1007/s43546-025-00931-7 |
| 72 | Manu et al. | 2024 | A state-of-the-art, systematic review of indoor environmental quality studies in work-from-home settings | Building and Environment | 10.1016/j.buildenv.2024.111652 |
| 73 | Mascarenhas et al. | 2025 | Tell me where and when you work, and I'll tell you how you recover: a systematic review of telework and non-standard working schedules as predictors of employee recovery | Scandinavian Journal of Work and Organizational Psychology | 10.16993/sjwop.353 |
| 74 | Mayer et al. | 2023 | Remote working and well-being: a review | SA Journal of Human Resource Management | 10.4102/sajhrm.v21i0.2009 |
| 75 | Meng et al. | 2025 | White-collar workers in the post-pandemic era: a review of risk and protective | Behavioral Sciences | 10.3390/bs15101313 |

| No. | Authors | Year | Title | Journal / Source | DOI |
|---|---|---|---|---|---|
| | | | factors for mental well-being | | |
| 76 | Miyake et al. | 2022 | Job stress and loneliness among desk workers during the COVID-19 pandemic in Japan: focus on remote working | Environmental Health and Preventive Medicine | 10.1265/ehpm.22-00005 |
| 77 | Morán et al. | 2022 | Impact of teleworking on the health and well-being of Peruvian workers in times of pandemic | Sustainability | 10.3390/su142315876 |
| 78 | Mutiganda et al. | 2022 | A systematic review of the research on telework and organizational economic performance indicators | Frontiers in Psychology | 10.3389/fpsyg.2022.1035310 |
| 79 | Navas-Martín et al. | 2026 | Working from home and indoor environmental quality: a scoping review | Applied Sciences | 10.3390/app16010250 |
| 80 | Ng | 2010 | Teleworker's home office: an extension of corporate office? | Facilities | 10.1108/02632771011023113 |
| 81 | Nguyen | 2021 | Factors influencing home-based telework in Hanoi (Vietnam) during and after the COVID-19 era | Transportation | 10.1007/s11116-021-10169-5 |
| 82 | Oda et al. | 2026 | Health of teleworkers: a scoping review on the assessment of the work-from-home environment | Journal of Occupational Health | 10.1093/joccuh/uiag007 |
| 83 | Onnis et al. | 2025 | Creating supportive technology-enhanced remote work environments: a review of the literature | Personnel Review | 10.1108/PR-05-2025-0547 |
| 84 | Ono et al. | 2023 | A scoping review of eye-tracking metrics as an indicator of negative mental health-related outcomes and its | Environmental and Occupational Health Practice | 10.1539/eohp.2023-0006-RA |

| No. | Authors | Year | Title | Journal / Source | DOI |
|---|---|---|---|---|---|
| | | | possible applicability in remote work situations | | |
| 85 | Palumbo et al. | 2023 | Looking for meanings at work: unraveling the implications of smart working on organizational meaningfulness | International Journal of Organizational Analysis | 10.1108/IJOA-09-2021-2980 |
| 86 | Parra et al. | 2022 | Towards an understanding of remote work exhaustion: a study on the effects of individuals' big five personality traits | Journal of Business Research | 10.1016/j.jbusres.2022.06.013 |
| 87 | Ploszaj et al. | 2025 | The relationship between remote work and job satisfaction: a literature review | BAR – Brazilian Administration Review | 10.1590/1807-7692bar2025250059 |
| 88 | Polspoel et al. | 2025 | Comparison of physical activity and sedentary behavior between telework and office work in a working population during the COVID-19 pandemic: a systematic review and meta-analysis of observational studies | BMC Public Health | 10.1186/s12889-025-22948-1 |
| 89 | Pyöriä | 2011 | Managing telework: risks, fears and rules | Management Research Review | 10.1108/01409171111117843 |
| 90 | Ríos Villacorta et al. | 2025 | Telework for a sustainable future: systematic review of its contribution to global corporate sustainability (2020–2024) | Sustainability | 10.3390/su17135737 |
| 91 | Rotimi et al. | 2024 | The potential challenges and limitations of implementing modern office design features in residential spaces: a SPAR-4-SLR approach | Buildings | 10.3390/buildings14103037 |

| No. | Authors | Year | Title | Journal / Source | DOI |
|---|---|---|---|---|---|
| 92 | Santos et al. | 2021 | Association between musculoskeletal pain and telework in the context of the COVID-19 pandemic: an integrative review | Revista Brasileira de Medicina do Trabalho | 10.47626/1679-4435-2021-812 |
| 93 | Santurtún and Shaman | 2023 | Work accidents, climate change and COVID-19 | Science of the Total Environment | 10.1016/j.scitotenv.2023.162129 |
| 94 | Schöne et al. | 2025 | The impact of working from home on sedentary behaviour and physical activity compared to onsite work in the working population: a systematic review and meta-analysis | BMC Public Health | 10.1186/s12889-025-24960-x |
| 95 | Seo and Kim | 2023 | Flexible work systems: preparing employees for the new normal | Journal of Business Strategy | 10.1108/JBS-03-2022-0054 |
| 96 | Shaholli et al. | 2024 | Teleworking and mental well-being: a systematic review on health effects and preventive measures | Sustainability | 10.3390/su16188278 |
| 97 | Siddique et al. | 2025 | A systematic review on work from home (WFH) and libraries: service delivery, tools used, challenges and opportunities | Global Knowledge, Memory and Communication | 10.1108/GKMC-04-2023-0152 |
| 98 | Šímová et al. | 2023 | Metaverse in the virtual workplace: who and what is driving the remote working research? A bibliometric study | Vision | 10.1177/09722629231168690 |
| 99 | Šímová et al. | 2024 | Metaverse in the virtual workplace | Vision | 10.1177/09722629231168690 |
| 100 | Staub | 2026 | Living and working on the move: individual-level insights into digital nomadism | Journal of Global Mobility | 10.1108/JGM-12-2025-0140 |

| No. | Authors | Year | Title | Journal / Source | DOI |
|---|---|---|---|---|---|
| 101 | Sun et al. | 2025 | Changing sense of place in hybrid work environments: a systematic review of place identity and employee well-being | Wellbeing, Space and Society | 10.1016/j.wss.2025.100236 |
| 102 | Tapasco-Alzate et al. | 2024 | Drivers of teleworker productivity: a systematic review of the empirical evidence | Communications in Science and Technology | 10.21924/cst.9.2.2024.1406 |
| 103 | Telu and Kumar | 2025 | Towards a sustainable future: a comprehensive review of employee well-being in hybrid work settings | Management and Sustainability | 10.1108/MSAR-10-2024-0182 |
| 104 | Ugemuge et al. | 2022 | Work from home – a growing trend in IT companies benefits, drawbacks, opportunities, and challenges | Journal of Datta Meghe Institute of Medical Sciences University | 10.4103/jdmimsu.jdmimsu_456_21 |
| 105 | Urien | 2023 | Teleworkability, preferences for telework, and well-being: a systematic review | Sustainability | 10.3390/su151310631 |
| 106 | Verma et al. | 2023 | The future of work post Covid-19: key perceived HR implications of hybrid workplaces in India | Journal of Management Development | 10.1108/JMD-11-2021-0304 |
| 107 | Vleeshouwers et al. | 2022 | The relationship between telework from home and the psychosocial work environment: a systematic review | International Archives of Occupational and Environmental Health | 10.1007/s00420-022-01901-4 |
| 108 | Wang and Hasan | 2026 | A systematic literature review: exploring the impact of inclusive leadership in hybrid teams | Cogent Business and Management | 10.1080/23311975.2026.2663579 |
| 109 | Wechsler et al. | 2024 | The impact of remote work using mobile information and communication technologies on physical | Ergonomics | 10.1080/00140139.2024.2304582 |

| No. | Authors | Year | Title | Journal / Source | DOI |
|---|---|---|---|---|---|
| | | | health: a systematic review | | |
| 110 | Wells et al. | 2023 | A systematic review of the impact of remote working referenced to the concept of work–life flow on physical and psychological health | Workplace Health and Safety | 10.1177/21650799231176397 |
| 111 | Willox et al. | 2023 | Benefits of individual preparation for team success: planning for virtual team communication, conflict resolution and belonging | Team Performance Management: An International Journal | 10.1108/TPM-05-2022-0040 |
| 112 | Wodajeneh et al. | 2023 | Ergonomic risk factors analysis in remote workplace | Theoretical Issues in Ergonomics Science | 10.1080/1463922X.2022.2157696 |
| 113 | Wütschert et al. | 2022 | A systematic review of working conditions and occupational health in home office | Work | 10.3233/WOR-205239 |
| 114 | Yousef | 2026 | Navigating global virtual teams: technological, cultural and leadership challenges in the era of remote work | Journal of Global Mobility | 10.1108/JGM-04-2025-0044 |

Note. Articles are listed alphabetically by first author. DOIs are given without the https://doi.org/ prefix.